\documentclass[useAMS,usenatbib]{mn2e}
\usepackage{graphicx}
\usepackage{color}
\newcommand{\hMsun}{$h^{-1}\rm{M_{\odot}}$}
\newcommand{\hMpc}{$h^{-1}\rm{Mpc}$}
\newcommand{\hGpc}{$h^{-1}\rm{Gpc}$}
\newcommand{\etal}{et al.~}

\newcommand{\Rvir}{R_\mathrm{vir}}
\newcommand{\Mvir}{M_\mathrm{vir}}

\newcommand{\Mth}{$M_\mathrm{th}$}
\newcommand{\MthA}{$M_\mathrm{th,1}$}
\newcommand{\MthB}{$M_\mathrm{th,2}$}
\newcommand{\Mdom}{$\mathrm{M}_\mathrm{dom}$}

\newcommand{\mnras}{MNRAS}
\newcommand{\apj}{ApJ}
\newcommand{\apjs}{ApJs}

\newcommand{\aj}{AJ}
\newcommand{\aap}{A\&A}
\newcommand{\prd}{Phys. Rev. D.}
\newcommand{\physrep}{Phy. Rep.}

\title[The cosmic mass density field]
{Halo based reconstruction of the cosmic mass density field}
\author[J. C. Mu\~noz-Cuartas, V. M\"{u}ller, J. E. Forero-Romero]
{J. C. Mu\~noz-Cuartas; V. M\"{u}ller; J. E. Forero-Romero \\
Astrophysikalisches Institut Potsdam, An der Sternwarte 16, 14482
Potsdam, Germany.}
\begin{document}

\date{Accepted XXXX December XX. Received XXXX December XX; in original form
  2011 February 28}

\pagerange{\pageref{firstpage}--\pageref{lastpage}} \pubyear{2002}

\maketitle

\label{firstpage}

\begin{abstract}
  
  We present the implementation of a halo based method for the
  reconstruction of the cosmic mass density field. The method employs
  the mass density distribution of dark matter haloes and its
  environments computed from cosmological N-body simulations and
  convolves it with a halo catalog to reconstruct the dark matter
  density field determined by the distribution of haloes. We applied
  the method to the group catalog of Yang \etal (2007) built from the
  SDSS Data Release 4. As result we obtain reconstructions of the
    cosmic mass density field that are independent on any explicit
    assumption of bias. We describe in detail the implementation of
  the method, present a detailed characterization of the reconstructed
  density field (mean mass density distribution, correlation function
  and counts in cells) and the results of the classification of large
  scale environments (filaments, voids, peaks and sheets) in our
  reconstruction. Applications of the method include morphological
  studies of the galaxy population on large scales and the realization
  of constrained simulations.
    
\end{abstract}



\section{Introduction}

Galaxy surveys have become a useful mean to investigate the large
scale structure of the Universe. Since late 60's with the Lick Galaxy
Catalog (Shane \& Wirtanen 1967), the large scale structure of the
Universe has been studied through the distribution of galaxies in
surveys like the CfA Redshift Survey (Huchra \etal 1983), 2MASS
Redshift Survey (Huchra 2000), the Las Campanas Redshift Survey
(Shectman \etal 1996), the 2dF survey (Colless \etal 2001) and the
Sloan Digital Sky Survey (York \etal 2000). These surveys provide us
with a wealth of data that have been used extensively in the study of
the properties and evolution of galaxies as well as in studies of the
properties of the mass distribution in the universe. Under the current
paradigm of structure formation ($\Lambda$CDM), galaxies are biased
tracers of the mass distribution, which is thought to be dominated by
dark matter. Nevertheless, observing the galaxy distribution is the
only way we have to study the large scale cosmic mass distribution.

From the theoretical point of view, the cosmic mass distribution can
be modeled as a smooth continuous function of the coordinates. From
that assumption, important physical insight can be obtained about the
process of structure formation and the dynamics of the
Universe. However the observed galaxies represent discrete objects,
therefore the discrete distribution of points has to be used to infer
the smooth continuous mass density field behind the observed galaxy
distribution. Such an investigation requires not only high amount of
good quality data, but also appropriate procedures able to reliably
reconstruct the smooth mass density field. These methods must be able
to deal with the inherent difficulties associated with the
observations like incompleteness, selection effects, geometrical
constraints and redshift space distortions.

Fortunately, current galaxy redshift surveys provide the required
amount and data quality to allow the reconstruction of the mass
density field. Previously, many efforts have been made to reconstruct
the density field using surveys like 1.2Jy IRAS redshift survey
(Fisher \etal 1995), 2dF Galaxy Redshift Survey (Erdo{\u g}du \etal
2004), 2MASS Redshift Survey (Erdo{\u g}du \etal 2006), the Luminous
Red Galaxy Sample in the SDSS (Reid \etal 2009), the 10k zCOSMOS
survey (Kova{\v c} \etal 2010) and the SDSS galaxy redshift survey
(Kitaura \etal 2009, Jasche \etal 2010). Almost of all of these works
have based the reconstruction on the distribution of individual
galaxies using the Wiener filter technique. Other methods have been
proposed, like Delaunay tessellation (Schaap \& van de Weygaert 2000),
Monge-Amp{\`e}re-Kantorovich method (Mohayaee \etal 2006), Bayesian
methods (Jasche \etal 2010) and halo based methods (Wang \etal 2009).

All of these methods use the galaxy distribution as tracers of the
cosmic mass distribution, from them, they compute the cosmic mass
distribution after smoothing the particle distribution on a grid and
convolving the particle distribution with a given smoothing
kernel. Since the reconstruction is based on the distribution of
galaxies, and they may be biased tracers of the mass distribution,
assumptions on the shape and functional dependence of the bias factor
are needed (Erdo{\u g}du \etal 2006, Mo \& White 1996).

Wang \etal (2009) have proposed a halo based reconstruction method
that uses dark matter haloes instead of galaxies as the point process
behind the reconstruction procedure, while it simultaneously enables
the inclusion of environments on the mass distribution around these
haloes. Although they only tested their procedure against simulations,
it seems to be quite promising, once one is willing to accept, first,
the validity of the results of N-body simulations as true realizations
of the large scale mass distribution, and second, assuming that from
the observations one can build a reliable halo catalog to make the
reconstruction. Currently both requirements seems to be fulfilled,
given the good performance of the cosmological simulations in the
concordance $\Lambda$CDM model to reproduce the features of the cosmic
large scale structure, and the current availability of data that makes
possible to build catalogs of groups of galaxies in the local Universe
(e.g. Tago \etal 2008, Tago \etal 2010, Yang \etal 2007, Crook \etal
2007, Berlind \etal 2006). Particularly Yang \etal (2007) provides a
group catalog based on the identification of groups of galaxies that
are located in the same dark matter halo. The realistic mass
assignment in these groups provides a unique opportunity to refer to
the dark matter haloes in the volume of the survey.

Reconstructions of the cosmic density field are necessary in the study
of the large scale structure through galaxy surveys. As was already
mentioned, they provide the continuous field that relates the
observations with the theoretical model describing our understanding
of the large scale structure. They are required to make calculations
of the intrinsic properties of the density field, like power spectrum
and mass variances (Jasche \etal 2010, Tegmark \etal
2004). Furthermore, they are used to extract information of the
cosmological parameters and study the dynamics of the Universe
(Percival \etal 2010). Usually cosmic environments are defined as a
function of the properties of the cosmic mass distribution
(Forero-Romero \etal 2009, Hahn \etal 2007). Identification and
classification of environments play a major role in the study of the
physical mechanisms involved in the process of galaxy formation and
evolution. High quality reconstructions of the local mass density
field coupled with reconstructions of the velocity field can offer an
interesting initial setup for the so called time machines that allow
the realization of constrained cosmological simulations that can be
used to study the time evolution of our local neighborhood (Kolatt
\etal 1996, Martinez-Vaquero \etal 2007, 2009, Gottl\"{o}ber \etal
2010, Lavaux \etal 2010b, Nuza \etal 2010).

In this work we address the problem of the reconstruction of the
cosmic mass density field following the halo based technique presented
in Wang \etal (2009). We extend the method to observational data and
apply it to the Fourth Data Release of the Sloan Digital Sky Survey
(SDSS-DR4) using the group catalog built by Yang \etal (2007) and the
outputs of cosmological N-body simulations. We perform a series of
tests to verify the quality and consistency of the
reconstruction and make use of the results to classify the morphology
of the large scale density field.

In this paper we first describe our methods, starting with the outline
of the approach. Then we describe the construction of the halo catalog
on which the reconstruction is based. Following, we show how to
compute the typical mass distribution in and around dark matter haloes
in simulations and to describe the procedures used to make the mass
assignment in the reconstruction, and finally we put all steps
together describing the procedure used to make the
reconstruction. Then we show our results, discuss the properties of
the reconstructed mass density field, compute different
characteristics of the density field, and as an application, we
perform the classification of the cosmic network from the
reconstruction.


\section{Methods} 
\label{sec:Methods}

\subsection{Outline}
Our reconstruction method follows closely the description presented in
Wang \etal (2009) and the basic ideas of the halo model (e.g. Cooray
\& Sheth 2002) where it is assumed that the cosmic mass is distributed
in haloes following a given mass density profile. We go beyond this
idea adding environment to the mass distribution around haloes
extending its associated mass distribution beyond the boundary of the
halo. First, let us suppose that the typical mass density distribution
in and around dark matter haloes of a given mass $M$, $\eta(r,M)$, is
a known function of the distance $r$ from the center of the halo and
the halo mass $M$. Let us assume also that the spatial coordinates and
masses of a set of dark matter haloes are known as some function
$\Phi(\vec{\bf r_i},M_i)$ where $\vec{\bf r_i}$ and $M_i$ are the
individual coordinates and masses for the ith halo. Then we define the
mass density field around the ith halo, $\rho_i(r_i,M_i)$, as the
convolution of the mass density distribution $\eta(r,M)$ with the set
of coordinates of each halo $\Phi(\vec{\bf r_i},M_i)$ as

\begin{equation}
  \rho_i(\vec{\bf r_i},M_i) = \eta(r,M) \otimes \Phi(\vec{\bf r_i},M_i) .
  \label{eq:convolution}
\end{equation}

With this way of defining the density distribution of individual
haloes, the total mass distribution $\rho(\vec{\bf r})$, arising from
the complete set of haloes is defined as

\begin{equation}
  \rho(\vec{\bf r}) = {\bf \hat{\Sigma}} ~\rho_i(\vec{\bf r_i},M_i) .
  \label{eq:reconstruction}
\end{equation}

\noindent
For well isolated haloes, for which the domains\footnote{As will be
  defined later, the domain of a halo is used in reference to the
  close volume or environment of the halo.} do not overlap each other,
the operator ${\bf \hat{\Sigma}}$ will represent a summation of the
density profiles at the different positions of the haloes. However
because haloes and domains overlap with each other, the operator ${\bf
  \hat{\Sigma}}$ has to be defined as the operation that composes the
mass density distribution in and around haloes depending on the
extension of the domain of the haloes and the environment where they
are located. ${\bf \hat{\Sigma}}$ will be defined operationally in
section \ref{sec:Recmeth}. Note that if $\eta(r,M) = k_w(\vec{\bf
  r}-\vec{\bf r_i})$, Eqs. \ref{eq:convolution} and
\ref{eq:reconstruction} will reduce to the typical way to compute the
smoothed density field with a kernel function $k_w$.

\begin{figure*}
  \includegraphics[width=17.0cm,angle=0]{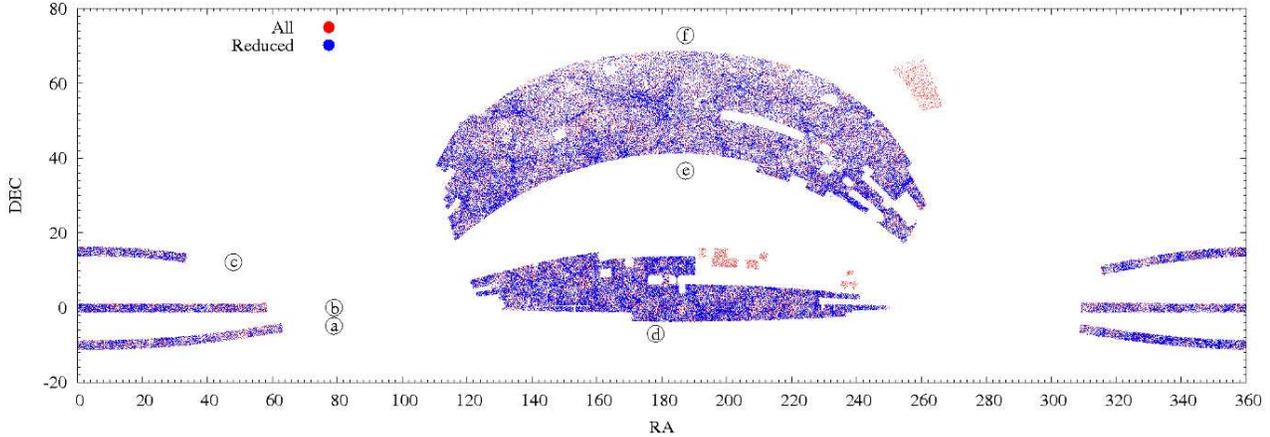}
  \caption{Geometry of the spatial distribution of groups in the group
    catalog (red) and the reduced sample used in this work (blue). The
    labels a, b, c, d and e are related to the labels of the regions
    shown in figure \ref{fig:reconstructions}.}
  \label{fig:SurveyGeometry}
\end{figure*}

The function $\Phi(\vec{\bf r_i},M_i)$ describing the positions and
masses of haloes is given by their coordinates and masses in the halo
catalog adopted for the reconstruction, while the typical mass
distribution in a halo of mass $M$ and its surroundings, $\eta(r,M)$,
is computed from cosmological simulations as will be described in the
next sections.

\subsection{The underlying group catalog and the construction of $\Phi(\vec{\bf r_i},M_i)$} 
\label{sec:Hcat}

One of the key ingredients in our approach is the catalog of
haloes. The use of a catalog of haloes allows us to trace the mass
distribution of well defined structures. Current redshift surveys like
2MASS and SDSS provide us with enough and accurate data to enable the
identification of groups of galaxies associated to dark matter haloes
with relatively low masses, down to $10^{11} - 10^{12}$ \hMsun~ (Crook
\etal 2007, Yang \etal 2007, Tago \etal 2005, 2010). This makes it
possible to work directly with the dark matter halo hosting these
groups of galaxies.

Here we define the concept of halo catalog and establish the
difference between it and a group catalog. We define a group catalog
as a collection of groups of galaxies that under clustering analysis
have shown to form an enhancement in the number density of galaxies
(Davis \etal 1985, Crook \etal 2007, Tago \etal 2008,Tago \etal 2010,
Berlind \etal 2006, Yang \etal 2006, Yang \etal 2007, Wen \etal
2009). On the other hand, a halo catalog is defined as a catalog with
positions and masses of dark matter haloes. Clearly one can build a
galaxy catalog from a halo catalog (i.e. using semi-analytic methods
in simulations) and the inverse, but the latter is a rather difficult
task since in the galaxy catalog one needs to relate galaxies hosted
in the same dark matter halo, to find the position of the halo and to
assign a mass.

For this work we decided to build the halo catalog from the catalog of
groups presented in Yang \etal (2007) prepared from the fourth data
release of the Sloan Digital Sky Survey, SDSS-DR4. We used the main
sample of the catalog (named as sample II in their paper) consisting
on 301237 groups in the redshift range from 0.01 to 0.2. The group
catalog is built from a sample of galaxies with $M^{0.1}_r - 5\log{h}
\leq -19.5$ in a way that restricts the estimated masses of haloes to
masses higher than $10^{11.6}$ \hMsun. Considering this constraint
and, in a compromise for a high resolution in the sampled mass in
haloes and a large volume of the survey to be reconstructed, we
decided to take a volume limited sample of haloes in the range of
redshifts between 0.01 and 0.1. To test for the effect of this choice,
we made a test reconstruction using a smaller sample volume, in
redshifts between 0.01 and 0.05 and found no changes in our results.

For the construction of the halo catalog, the group catalog of Yang
\etal (2007) provides an estimate of the masses of the associated
haloes. We assume the mass associated to each halo to be the one
estimated using the ranking of the halo luminosities, and when not
present, we used the one obtained from the ranking on the stellar mass
(see Yang \etal 2007 for a detailed description about these two mass
assignments).

The method of reconstruction explicitly depends on a mass threshold
establishing the minimum halo mass to be resolved. In our main
reconstruction we have chosen a value for the mass threshold
\MthA=$10^{11.5}$\hMsun. As will be shown later in Sections
\ref{sec:sims} and \ref{sec:ResultsGeneral}, to test for the influence
of the value of \Mth, we have made a second reconstruction of the
density field using a mass threshold \MthB=$10^{12.7}$\hMsun.

We assume that the mass assigned to haloes in the catalog corresponds
to the virial mass. Although this might be a strong assumption, we
expect the differences between the assumed and the true virial masses
not to be larger than a factor of 2 (White 2001). In what follows, all
references to halo masses will be implicitly assumed to correspond to
$\Mvir$, unless it is explicitly stated.

Furthermore, the geometry of the group catalog can be quite complex,
introducing difficulties in the process of the reconstruction and
during the analysis of the density field. In order to simplify our
procedures associated with the complex geometry of the survey, we have
made a reduction in the geometry of the halo catalog and worked with
groups that are contained inside a compact geometry, removing small
slabs or peaks in the border of the window survey. Figure
\ref{fig:SurveyGeometry} shows in equatorial coordinates the original
and final distribution of groups after the geometry reductions.

All this selection criteria applied to the group catalog leaves us
with a set of 86128 haloes in the halo catalog for the reconstruction
with \MthA~ and 8165 haloes for \MthB. Each one of these sets of haloes
determine the function $\Phi(\vec{\bf r_i},M_i)$ necessary for the
reconstruction.

Assuming that the luminosity-weighted redshift of each group ($z_i$)
corresponds to the redshift of each halo in the halo catalog, we
compute the comoving cartesian coordinates for each halo from the
coordinates of the associated groups as
\begin{eqnarray}
x_i &=& r_i\cos(\delta_i)\cos(\alpha_i), \nonumber\\
y_i &=& r_i\cos(\delta_i)\sin(\alpha_i), \\
z_i &=& r_i\sin(\delta_i), \nonumber
\end{eqnarray}

\noindent
where $\alpha_i$ and $\delta_i$ are the right ascension and declination
of each galaxy group in the group catalog and $r_i$ is the comoving
distance of the respective halo, given by

\begin{equation}
  r_i = c \int_0^{z_i}\frac{dz}{H_0\sqrt{\Omega_m(1+z)^3 + \Omega_{\Lambda}}},
\end{equation}

\noindent
with $c$ being the speed of light, $\Omega_m=0.24$ and
$H_0=73~\rm{km/s~Mpc}^{-1}$ is the Hubble constant at present
time. Then, our sample of haloes is contained in a set of slices of
$\sim 403$ \hMpc~ radius in comoving units.

Finally, for each halo we computed its radius $\Rvir$ from their
masses using Eq. \ref{eq:Rvir} below, assuming that they all are
situated at $z=0$.

The reconstruction of the density field presented in this paper has
been done in redshift space, and therefore it must contain redshift
space distortions in the coordinates of the haloes coming from the
group catalog. However since the reconstruction method is based in
haloes, most of the highly nonlinear redshift effect distortions
should not be present, and only mildly linear regime effects may be
accounted for. As was shown in Wang \etal (2009), this method can be
used to make reconstructions of the density field in real space,
nevertheless we have decided to work in redshift space since computing
properly the effects of the large scale structure on the velocity
field requires an isotropically sampled volume. As the sampling
geometry of the survey used for the reconstruction (SDSS-DR4) is
highly anisotropic, consisting on several slabs, the large empty
regions in the volume of the reconstruction will make it unreliable to
estimate the influence of the unseen large scale structures on the
velocities of the objects in the reconstruction region.

\subsection{Numerical Simulations and the estimation of $\eta(r,M)$} 
\label{sec:sims}

\subsubsection{Halo and domain identification and halo properties}

\begin{figure}
  \includegraphics[width=7.4cm,angle=270]{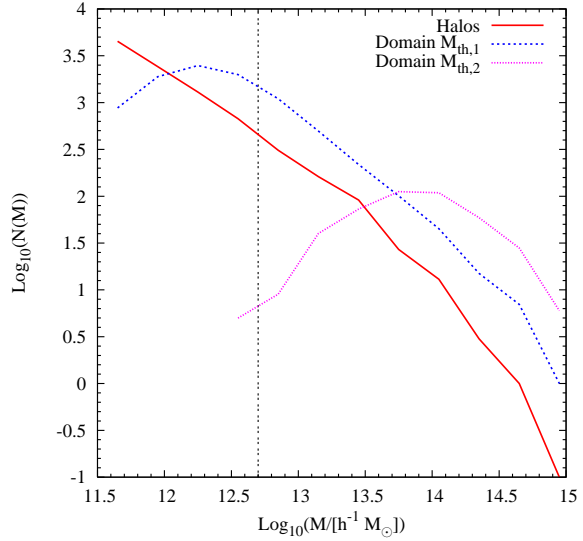}
  \caption{Halo mass function for the haloes and their domains in the
    simulation for two different mass thresholds \MthA=$10^{11.5}$ and
    \MthB=$10^{12.7}$\hMsun. The vertical line shows the cut on the
    halo mass function for the second mass threshold, \MthB.}
  \label{fig:Mfunctions}
\end{figure}

Our method makes use of the masses and positions of haloes taken from
the group catalog of the SDSS and convolves it with the mass
distribution in and around dark matter haloes estimated from
cosmological simulations. It combines results of simulations and
observations in a unique effort seeking for a realistic reproduction
of the cosmic mass distribution.

In this work we have used a cosmological simulation ran under the
standard spatially flat $\Lambda$CDM cosmology in a cubic box of
100\hMpc~ with $512^3$ particles and cosmological parameters given by
$\Omega_m=0.24$, $\Omega_{\Lambda}=0.76$, $\sigma_8=0.76$ and
$h=0.73$. In order to be fully consistent, these cosmological
parameters where intentionally chosen to be the same set of parameters
used in the construction of the halo catalog. No further corrections
for cosmology have to be considered, reducing in particular the
uncertainties in the mass assignment of haloes (see Yang \etal
2007). For the determination of the mass density profiles we used the
last snapshot of the simulation at $z=0$.

Haloes are identified using a FoF algorithm with a linking length of
0.2 times the mean interparticle distance. The minimum number of
particles per halo $N_{\rm min}$ is chosen according to the mass
threshold $M_{th}$ used to do the reconstruction as $N_{\rm
  min}=M_{\rm th}/m_p$, where $m_p$ is the mass of a particle in the
simulation. We fixed the mass threshold for the reconstruction \MthA~
to be $10^{11.5}$\hMsun, which results in a minimum number of 637
particles per halo. This value of $10^{11.5}$\hMsun~ is slightly
smaller than the $10^{11.6}$\hMsun~ minimum mass in the halo catalog,
and was defined in that way to be able to account for the typical mass
distribution of haloes at its low mass end. As can be seen in table
\ref{tab:RecPars} the value of $10^{11.6}$\hMsun~ is almost at the
center of the first mass bin, therefore, no major effect is expected
from this choice. This will be also verified later, when we will show
the independence of the general properties of the reconstruction on
\Mth.

For each halo we define its center as the position of the particle
with the lowest potential energy. The virial mass of the halo is
defined as the mass of the sphere enclosing a mean density equal to
$\Delta_{\mathrm{vir}}\rho_{\mathrm{crit}}$, where
$\Delta_{\mathrm{vir}}$ is the virial density contrast computed using
the fitting formula from Brian \& Norman (1998) and
$\rho_{\mathrm{crit}}$ is the critical density of the Universe. For
$z=0$ and the cosmology assumed for this work
$\Delta_{\mathrm{vir}}=92.8$. The virial radius $R_{\mathrm{vir}}$ of
the haloes is computed as the radius at which the enclosed mass equals
$\Mvir$, and is related to this by

\begin{equation}
  \Rvir=\left(\frac{3\Mvir}{4\pi \Delta_\mathrm{vir}\rho_\mathrm{crit}} \right)^{1/3}.
  \label{eq:Rvir}
\end{equation}

The domain of a halo represents the unique region of the Universe that
is closest to the halo than to any other. Following Wang \etal (2009)
we identify the domain of haloes in the simulation using the particle
distribution. The domain of a halo $h$ is identified as the region of
space containing all particles for which the weighted distance measure
$d_{hj}/R^h_{\mathrm{vir}}$ between the halo $h$ and the particle $j$
is minimal across all the population of haloes. It is implicit here
that the particle $j$ is not bound to any other halo by the FoF
procedure and $d_{hj}$ is the distance between the halo of virial
radius $R^h_{\mathrm{vir}}$ and the particle.

Figure \ref{fig:Mfunctions} shows the mass function of haloes and
their domains for our simulation at the two different mass thresholds
\MthA=$10^{11.5}$ and \MthB=$10^{12.7}$ \hMsun. Note that for \MthA~
both lines (haloes and domains) are almost parallel for all masses
above $10^{12.7}$\hMsun, with discrepancies at the low and high mass
end. It shows that very massive haloes tend to have more extended and
massive domains, while at the low mass end it is clear that low mass
haloes may have very small domains or can even have no domain at
all. The mass function for \MthB~ shows a higher number of haloes with
massive domains. This can be easily interpreted, since using a higher
mass threshold implies that less structures are resolved as haloes and
therefore more mass is associated to the domain of more massive
haloes.

\begin{figure*}
  \includegraphics[width=15.0cm,angle=270]{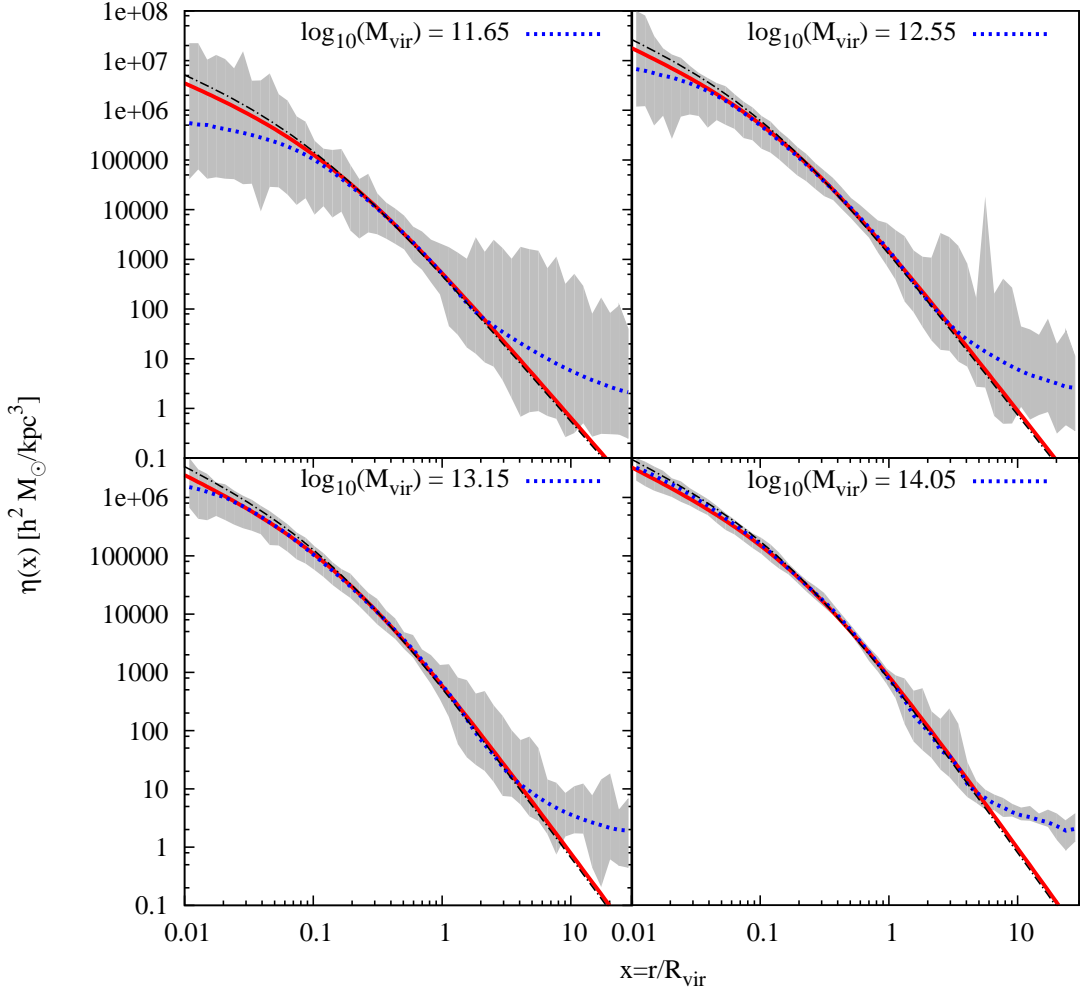}
  \caption{Mean mass density distribution for haloes in four different
    populations. In each panel, the blue dotted line shows the mean
    mass density distribution $\eta(r,M)$, while the shadow region
    shows the scatter of individual density profiles of haloes
    computed from Eq.  \ref{eq:Densprofile} with $N_h=1$. The red
    solid line shows the associated NFW density profile for each mass
    for a WMAP3 cosmology (Macci{\`o} \etal 2008), while the black
    dot-dashed line shows the same for a WMAP5 cosmology
    (Mu\~noz-Cuartas \etal 2010).}
  \label{fig:Dprofiles}
\end{figure*}

\subsubsection{Computing the mean mass density distribution of haloes $\eta(r,M)$}
\label{sec:etar}

Once the virial masses, radii and domains of haloes are determined, we
proceed in estimating the mass density distribution associated to each
halo. For this step, we bin the haloes in the simulation by mass, each
of these mass bins will be called a population of haloes. Then we
compute the mean mass distribution $\eta(r,M)$ for each population
from the simulation. See table \ref{tab:RecPars} for a detailed
description of the mass binning used in this work.

Assuming that each particle in the simulation box has an associated
volume $V_p$, we computed $\eta(x,M)$ binning the particle
distribution in radial logarithmic bins of width $dx$ where
$x=r/R_{\mathrm{vir}}$ is the normalized radius at distance $r$ from
the center of the halo. For a given population with $N_h$ haloes we
computed the mean mass density as the ratio of the total mass enclosed
in that radial bin from all haloes in the population divided by the
volumes of the $N_p$ particles per halo inside the radial bin of width
$dx$, explicitly

\begin{equation}
  \eta(r,M) = \frac{\sum^{Nh}_h\sum^{Np}_pm_p}{\sum^{Nh}_h\sum^{Np}_pV_p}.
 \label{eq:Densprofile}
\end{equation}

The volume of every particle $V_p$ has been computed using a Delaunay
tessellation over the particles in the simulation box (O'Rourke 1998),
where the Delaunay tessellation has been computed using the freely
available library Qhull (Bradford \etal 1996). As it has been noted in
Wang \etal (2009), for a given population, $\eta(r,M)$ is nothing else
than the correlation function between haloes and the mass distribution
around them, so the reconstruction method carries implicitly the
information of the correlation function of the simulation (and
therefore, of the cosmology that has been adopted) through
$\eta(r,M)$.

Figure \ref{fig:Dprofiles} shows the typical density profiles
$\eta(r,M)$ estimated for four different halo populations, compared
with the individual density profiles for individual haloes in that
population (i.e. we use Eq. \ref{eq:Densprofile} for each halo
separately) and with the standard NFW profile with a mean
concentration parameter as given in Macci{\`o} \etal (2008) for a
WMAP3 and Mu\~noz-Cuartas \etal (2010) for a WMAP5 cosmology.

In the figure the grey region shows the scattered distribution for the
individual density profiles of haloes. As it was expected, the scatter
is larger for the low mass bins. At $r<\Rvir$ the scatter is high
since low mass haloes have a high scatter of concentration
parameters. At $r>\Rvir$ the scatter is large due to the many
different environments in which the low mass haloes are located. We
emphasize that this scatter is not due to resolution effects, since
the less massive haloes in this work have been identified with at
least 637 particles, this is a high enough number of particles to
resolve the density profile (Mu\~noz-Cuartas \etal 2010, Trenti \etal
2010). Note also in the figure that the mean density profiles
$\eta(r,M)$ approach satisfactorily the NFW shape, even for
$r>\Rvir$. The comparison of $\eta(r,M)$ with the mean density profile
for the WMAP3 and WMAP5 cosmologies shows that in principle one could
relax the requirement that the simulations used to get $\eta(r,M)$
must have exactly the same cosmology as the one used to create the
halo catalog.

\begin{figure}
  \includegraphics[width=7.6cm,angle=270]{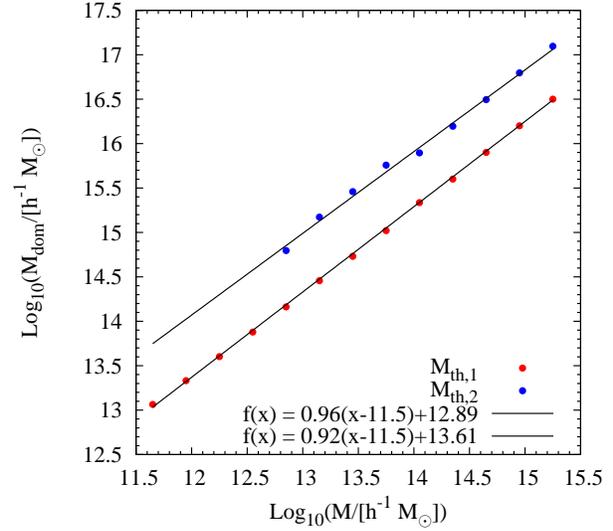}
  \caption{Initial mass assignment in domains of haloes in populations
    with mean mass $M$. The two sets of points represent the initial
    mass in domains of haloes in a population with mean mass $M$
    computed from Eq. \ref{eq:Mdomain} for two different values of the
    mass threshold, \MthA$=10^{11.5}$ and \MthB$=10^{12.7}$ \hMsun.}
  \label{fig:Massignement}
\end{figure}

\subsubsection{Mass assignment in domains}
\label{sec:Massignment}
Now we describe the procedure used to assign masses in the domain of
each population of haloes. As will be shown in the next section,
haloes may start with an initial guess for its domain mass, but during
the reconstruction procedure, depending on the environment where they
are located, this initial mass is reduced until a final mass is
assigned to the domain of each halo. We use the term ``initial mass in
domains" in reference to the guessed initial value for the mass in the
domain, and ``final mass in domains" in reference to the real final
mass associated with the domain of the halo after the reconstruction
is completed.

Since each halo in the simulation has been assigned to a population,
all of the haloes in the same halo population will be characterized by
the same mean mass density distribution, and in particular, the same
initial mass in their domains. The initial mass in each domain of
haloes is computed from the mean density profile $\eta(r,M)$ as

\begin{equation}
  M_\mathrm{dom}(M) = 4\pi \int^{\mathrm{r_{dom}}}_{\Rvir} \eta(r,M) r^2 dr,
  \label{eq:Mdomain}
\end{equation}

\noindent
where for $\eta(r,M)$ we have used the numerical values obtained from
the simulation for each halo population and $r_{\mathrm dom}$ was set
to $30\Rvir$. For each population $\Rvir$ was calculated consistently
with the value of the mean mass of that population using
Eq. \ref{eq:Rvir}.

Figure \ref{fig:Massignement} shows how the mass assignment procedure
relates the mean mass of the halo population with its mass in the
domain as a power law $M_\mathrm{dom}(M) = A M^{\alpha}$. The
normalization constant $A$ is close to $10^2$ and $10^3$ for
$M_\mathrm{th}=10^{11.5}$ and $M_\mathrm{th}=10^{12.7}$ \hMsun,
respectively, showing that domains of perfectly isolated haloes could
be more or less one hundred or one thousand of times more massive than
the haloes themselves. This proportionality grows directly with the
value of $M_\mathrm{th}$. It is worth to note that this factor of
about ten difference in the normalizations is close to the mass ratio
between the two different mass thresholds we choose. This difference
shows that for a given \Mth~ the mass we ignore in the form of dark
matter haloes goes directly to the domains in the form of cosmic
background mass.

\subsection{Reconstruction} 
\label{sec:Recmeth}

\begin{figure*}
  \begin{tabular}{cc}
    \includegraphics[width=7.4cm,angle=0]{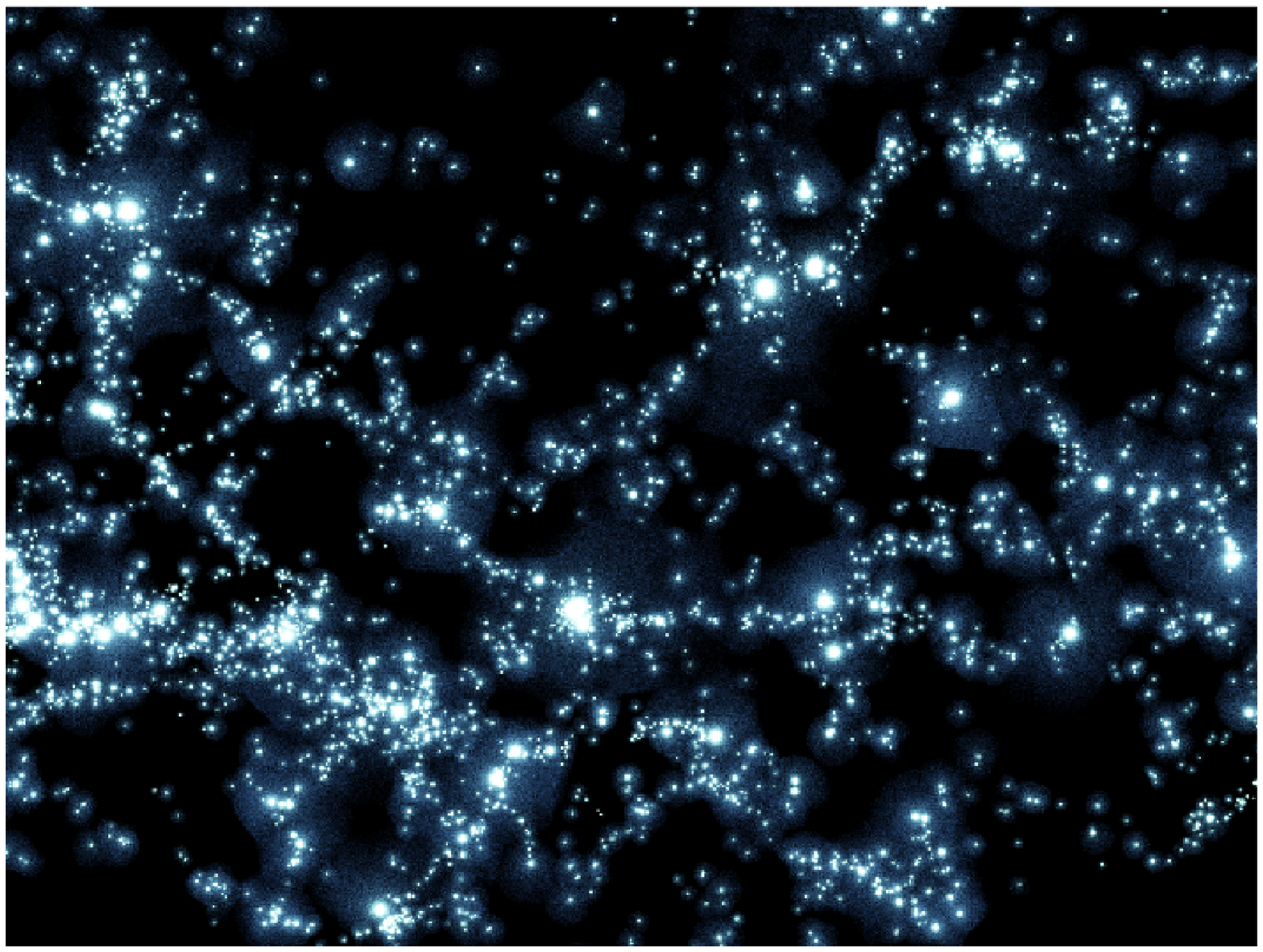}
    \includegraphics[width=7.4cm,angle=0]{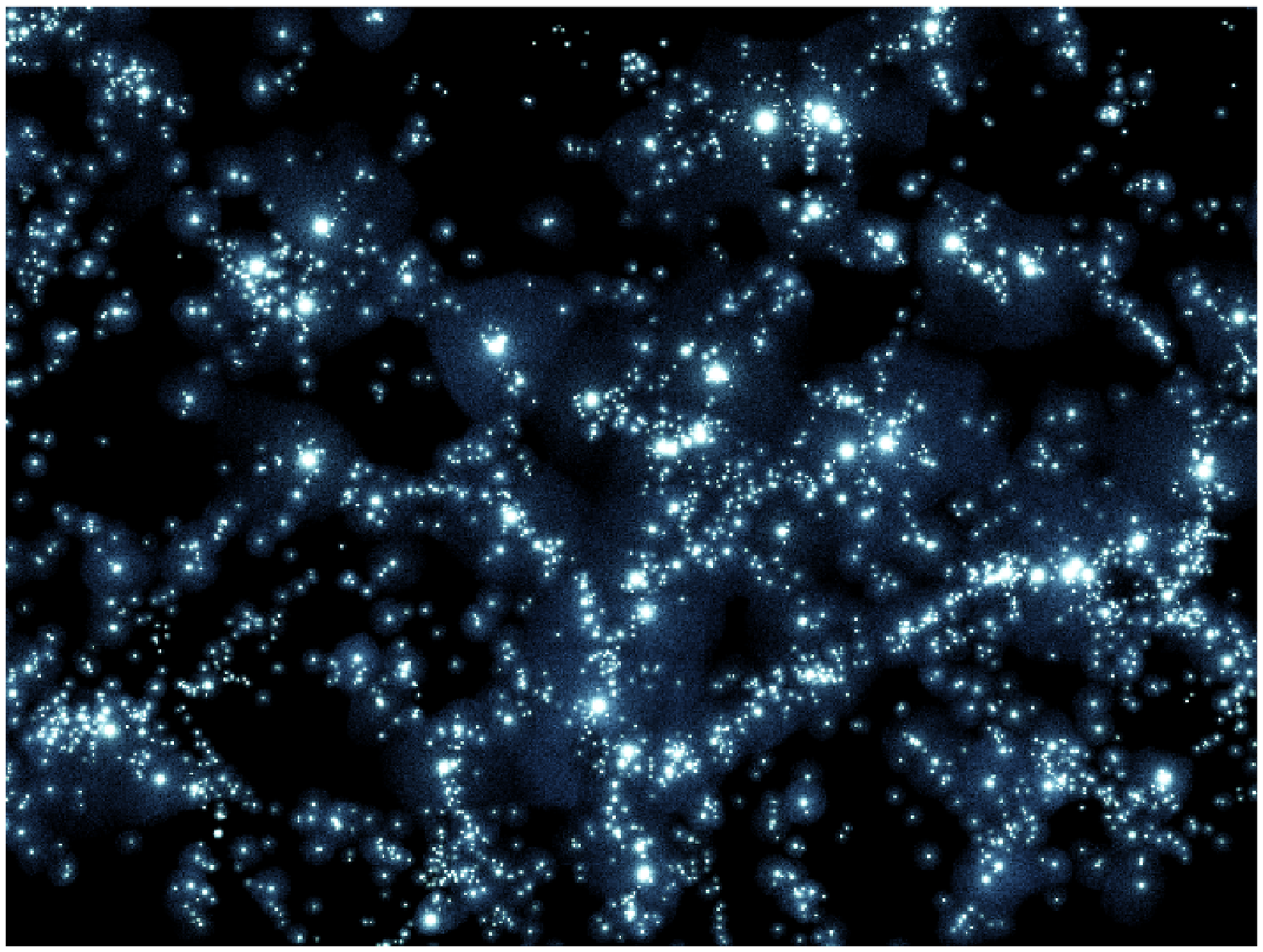}\\
    {\bf (a)} \hspace{7cm} {\bf (b)}\\
    \includegraphics[width=7.4cm,angle=0]{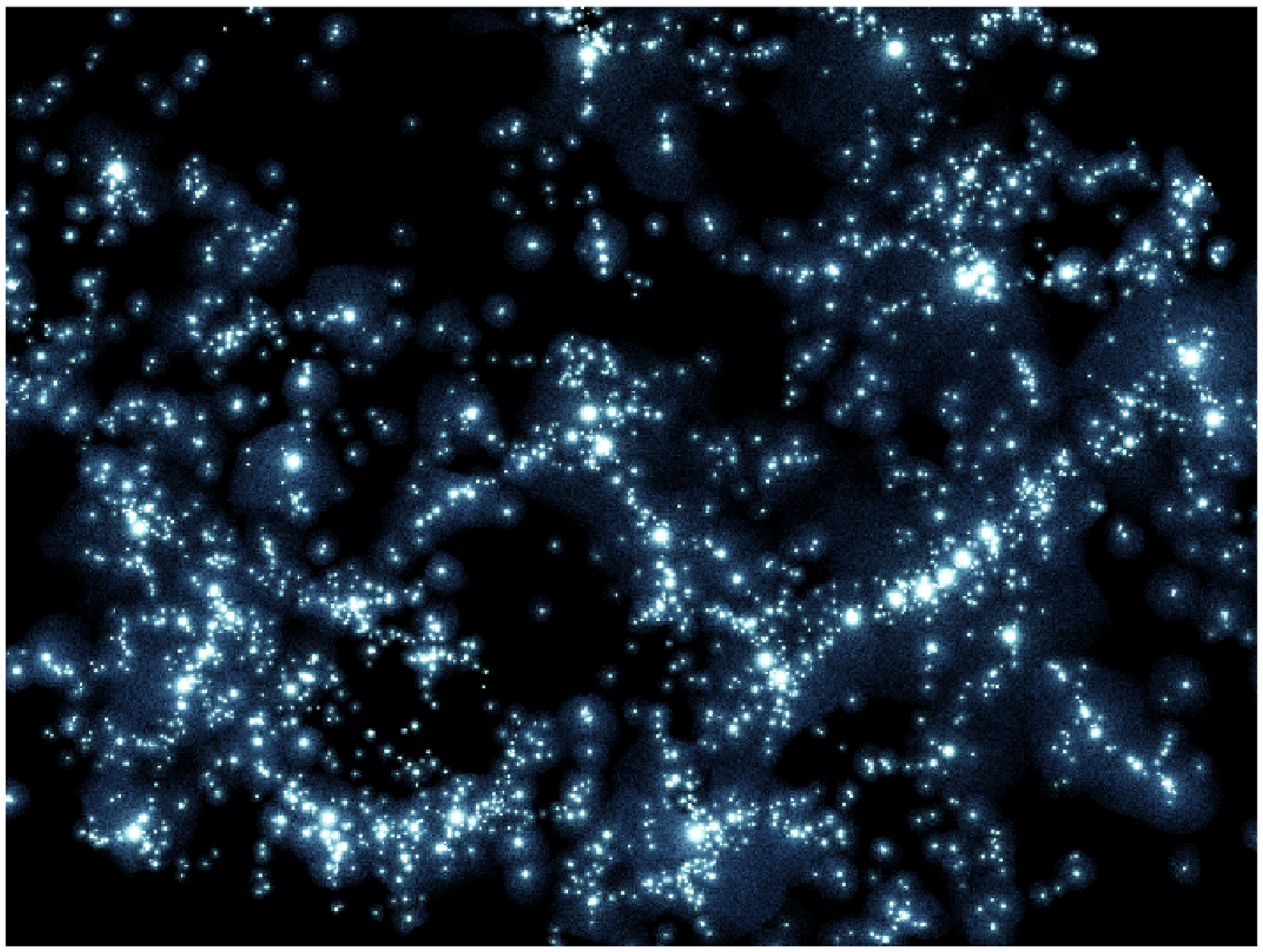}
    \includegraphics[width=7.4cm,angle=0]{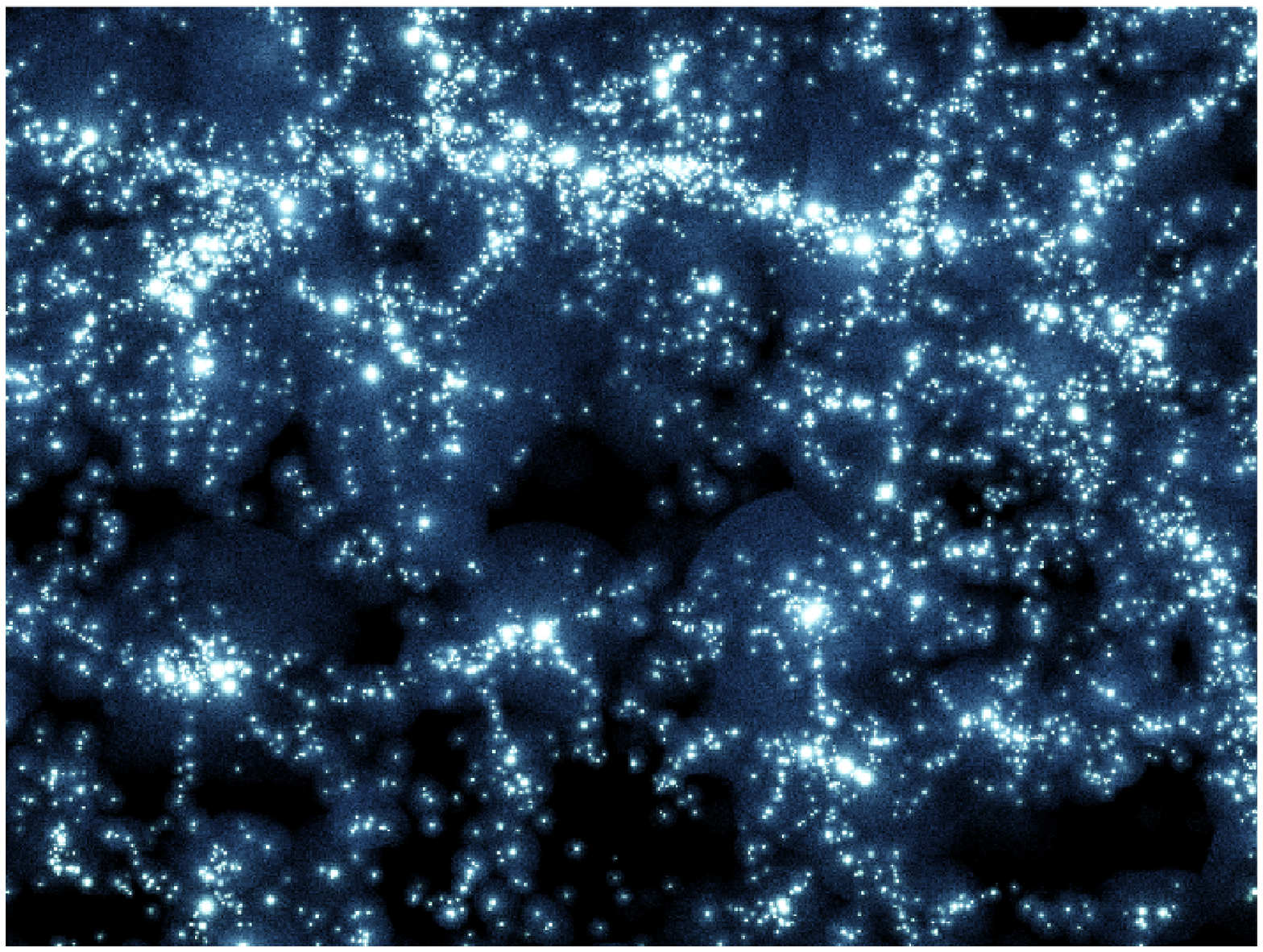}\\
    {\bf (c)} \hspace{7cm} {\bf (d)}\\
    \includegraphics[width=7.4cm,angle=0]{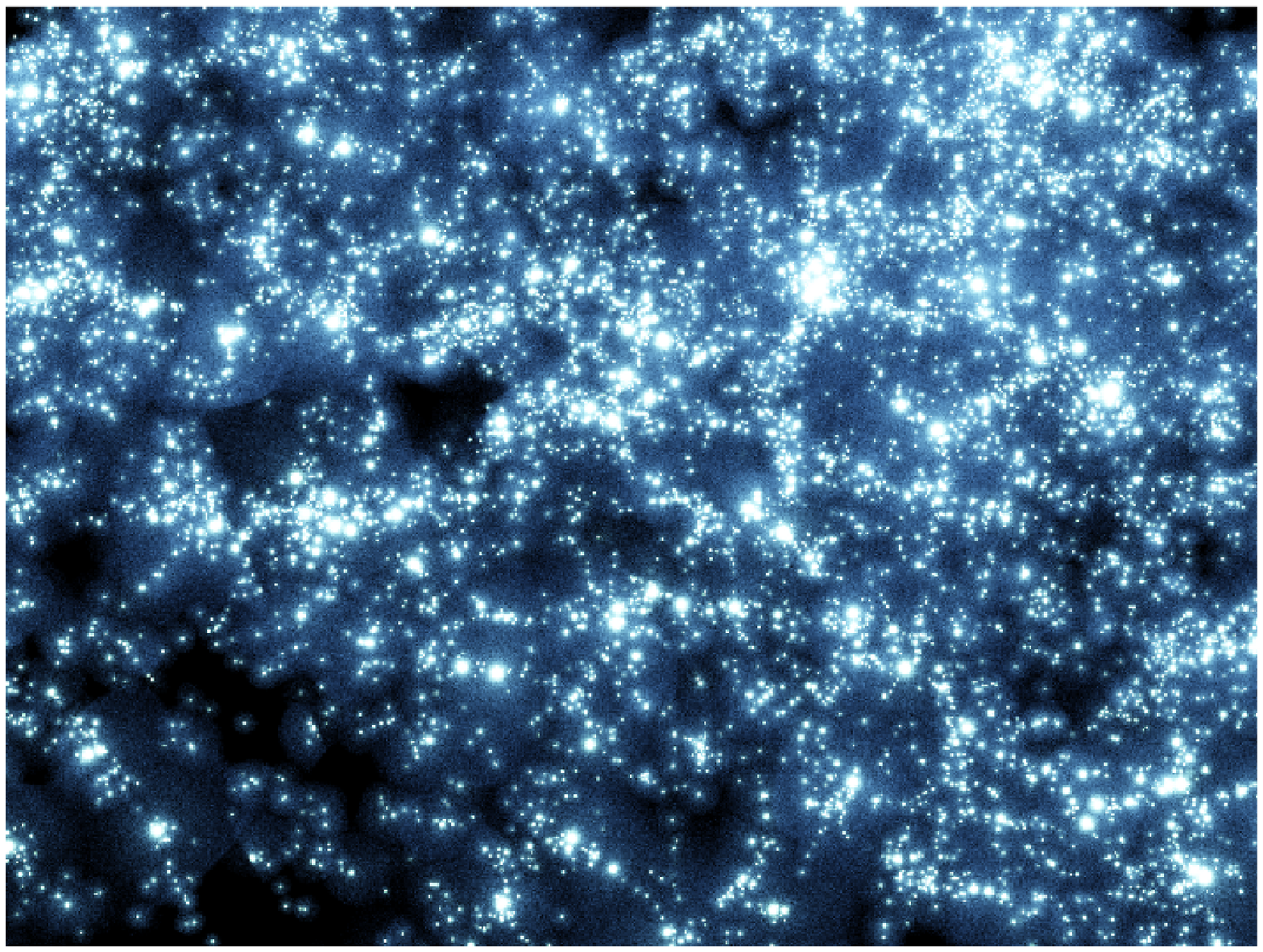}
    \includegraphics[width=7.4cm,angle=0]{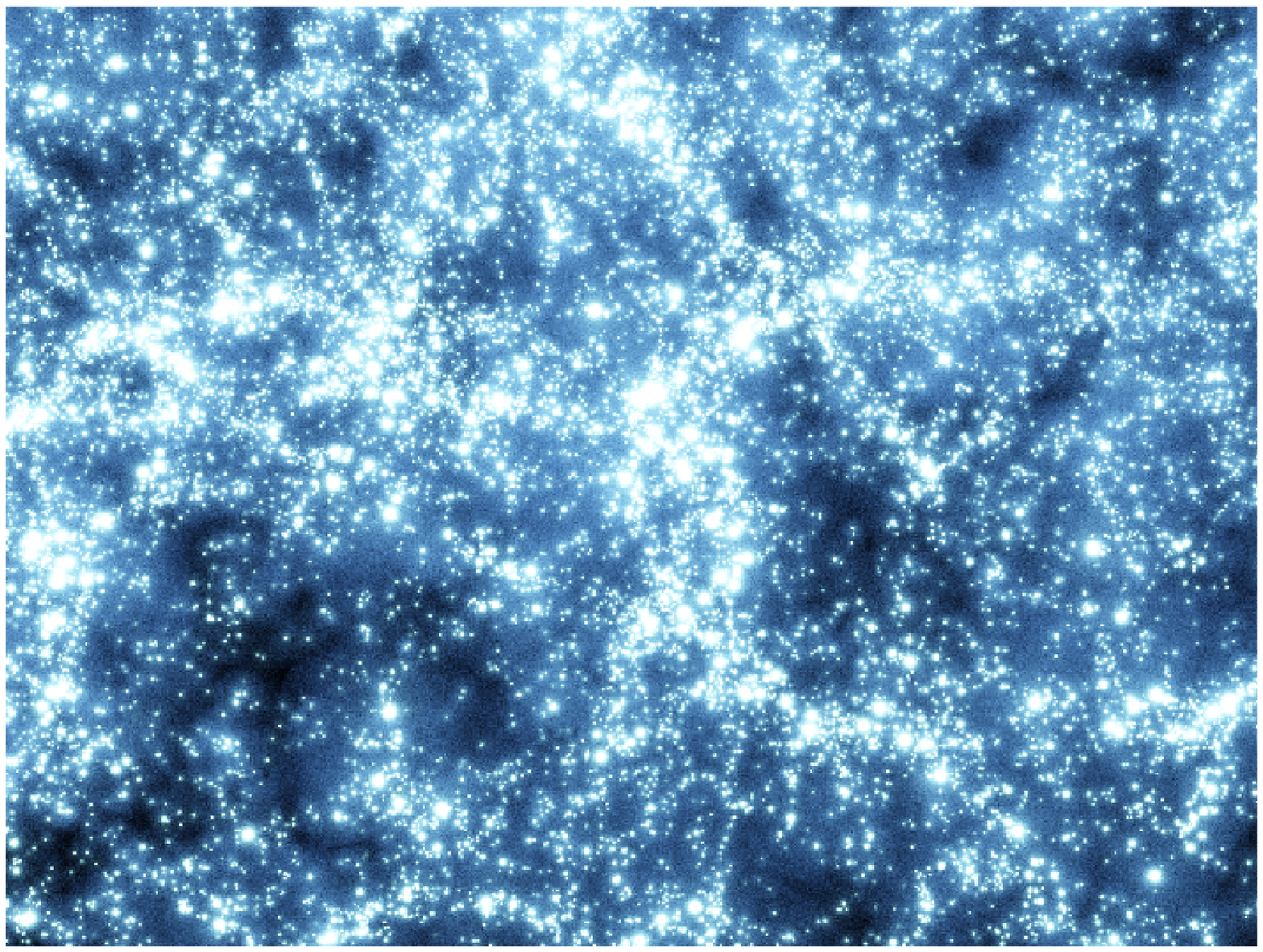}\\
    {\bf (e)} \hspace{7cm} {\bf (f)}\\
      
  \end{tabular}
  \caption{Visual appearance of the reconstruction for the
    different slabs of the survey. Labels a, b, c, d, e and f are
    related to regions with the same label in figure
    \ref{fig:SurveyGeometry}. Panels a, b and c show the
    reconstruction for the three slices of the southern caps of the
    survey. Panel d shows a section 20\hMpc~ thick of the north cap
    and panels e and f show two wedges of the northern cap. Each
    region covers an area of $330\times250$\hMpc}
  \label{fig:reconstructions}
\end{figure*}

Having all of the ingredients, one can proceed to use
Eqs. \ref{eq:convolution} and \ref{eq:reconstruction} to make the
reconstruction of the density field using the function describing the
positions and masses of the set of haloes $\Phi(\vec{\bf r_i},M_i)$
from the halo catalog and assigning them to a halo population
according to their masses, this will provide the form of $\eta(r,M)$
and the initial mass in the domain of every halo.

In practice what we do for each halo in $\Phi(\vec{\bf r_i},M_i)$ is:

\begin{enumerate}

\item Assign haloes in $\Phi(\vec{\bf r_i},M_i)$ to a given halo
  population according to their masses, $M_i$

\item Assign an initial mass to the domain of the halo according to
  its population and assign a total number of sampling particles. To
  obtain a final reconstruction consistent with the properties of the
  halo catalog, during the reconstruction, we force the halo to have
  the same mass inside the virial radius as in the catalog, so the
  total number of sampling particles is divided in two sets, halo
  particles $N_h=\Mvir/m_\mathrm{rec}$ and domain particles
  $N_d=M_\mathrm{dom}/m_\mathrm{rec}$ for each halo of mass $\Mvir$
  and domain mass $M_\mathrm{dom}$, and where $m_\mathrm{rec}$ is the
  mass of each sampling particle in the reconstruction.

\item Then we generate a particle realization of the density profile
  $\eta(r,M)$ at the position of the halo for the $N_h$ particles
  inside the virial radius ($r \leq \Rvir$). Once the halo has been
  realized, we proceed to generate a particle realization of the
  density profile $\eta(r,M)$ in the domain of the halo ($r>\Rvir$)
  with $N_d$ particles. In this step, only particles that fall inside
  the domain of the halo, computed across the complete halo catalog
  $\Phi(\vec{\bf r_i},M_i)$, are accepted to be part of the
  reconstruction. This step defines the operator ${\bf \hat{\Sigma}}$
  introduced in Eq. \ref{eq:reconstruction}. Here one can see that
  because of the rejection of the particles outside of the domain of
  the current halo, the operation ${\bf \hat{\Sigma}}$ can not be
  interpreted formally as a convolution and can only be defined
  operationally as we just did. Note that, as was already mentioned,
  in this step the actual number of particles in the domain $N_d$ and
  its mass $M_\mathrm{dom}$ will be different (lower) than the values
  assumed initially.

\end{enumerate}

An issue to consider during the procedure of reconstruction is related
to the border and the geometry of the survey. Although we have made
simplifications to the geometry of the distribution of haloes, the
SDSS-DR4 and therefore the halo catalog has a geometry that makes it
difficult to implement the method. In particular, haloes in the border
of the survey have large fractions of totally empty space around them
due to the boundary. For these haloes, during the reconstruction, the
extension of their domains in that direction will be larger than
expected if the unobserved haloes where present in the data in that
region. This will represent unphysically the mass distribution in
those regions. To avoid these drawbacks we use an auxiliary catalog of
haloes filling the empty regions of the survey and forming a thick
envelope around the different slabs of the survey. This envelope was
built taking haloes from a simulation of box size 1\hGpc. These haloes
are used not to build the reconstruction of the density field around
them, but for constraining the extension of the domain of the haloes
in the borders of the survey.


\section{Results}
\label{sec:results}

\subsection{General properties}
\label{sec:ResultsGeneral}


\begin{table}
\begin{center}
\begin{tabular}{|c|c|c|}\hline \hline

Name      & \MthA & \MthB \\\hline

$N_{\mathrm bins}$     & 13    & 9    \\
$\log_{10}$(\Mth/\hMsun) & 11.5  & 12.7 \\
$M_{\rm bin}$    & 11.65,..., 0.3dex & 12.85,..., 0.3dex \\
$N_{\mathrm haloes}$  & 86128 & 8165 \\
$m_{\mathrm{rec}}$     & 0.496 & 0.496 \\
Total $N_{\mathrm p}$  & $\sim2.3\times10^8$ & $\sim2.3\times10^8$ \\
\hline
\hline
\end{tabular}
\end{center}
\caption{Parameters used for the reconstruction of the density
  field. $N_{\mathrm bins}$ is the number of mass bins used to make
  the reconstruction, it also corresponds to the number of halo
  populations defined to compute $\eta(r,M)$. $\rm{M}_{\rm bin}$ is a
  detail on the mass binning, from the lower minimum mean mass in the
  bin until the last one using bins of width
  0.3dex. $N_{\mathrm{haloes}}$ represents the number of haloes used
  to make the reconstruction (the number of haloes in $\Phi(\vec{\bf
    r_i},M_i)$). $m_{\mathrm{rec}}$ is the particle mass in units of
  $10^{10}$\hMsun. Total $N_{\mathrm p}$ is the total number of
  particles in the reconstruction.}
  \label{tab:RecPars}
\end{table}


\begin{figure}
  \includegraphics[width=8.0cm,angle=270]{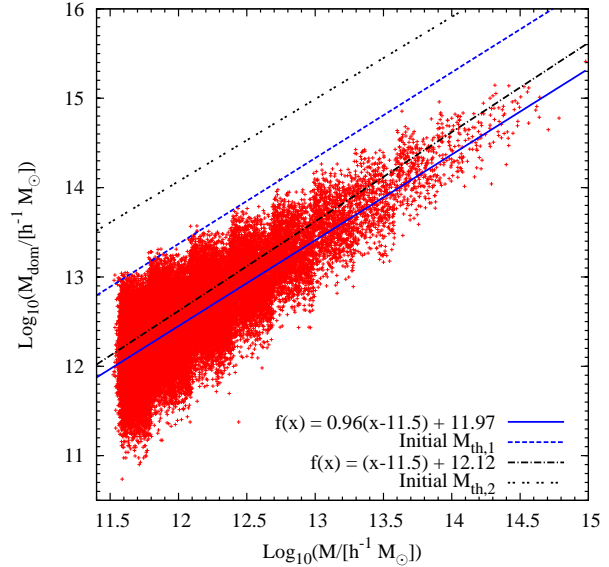}
  \caption{Halo - domain mass relation for the reconstructions of the
    density field. Points show the scatter of the domain masses for
    the haloes in the reconstruction with \MthA. The solid blue
    line shows a linear fit to the data. The dot-dashed black line
    represents the relation for the realization with \MthB~ and the
    dotted and dashed lines on top show the initial mass assignment
    already shown in Figure \ref{fig:Massignement}.}
  \label{fig:MdomMvir}
\end{figure}

The outcome of the reconstruction procedure just described is a set of
particles tracers of the density field distributed according to the
mass distribution in haloes and their domains. The resulting particle
distribution resembles the particle distribution in N-body
simulations. Their positions are constrained by three factors: the
positions of the haloes in the halo catalog, their environments and
the shape of the mass distribution around halos of a given mass.

Figure \ref{fig:reconstructions} shows slices of the reconstruction
for the different slabs on the survey, as described in the
caption. After a visual inspection it can be clearly seen how the
reconstruction method not only allows for the identification of high
density regions, but also clearly allows the identification of empty
regions. The slices were chosen to be thick enough ($\geq$ 10\hMpc) to
enclose a representative number of haloes and to make sure that the
structures (filaments or voids) observed are not due to geometrical
incompleteness of the survey. In panel (d) one can also see a large
structure running almost at the same distance from the origin across
all of the slice, that corresponds to the Great Wall, and just below
it one can see a huge empty region (Gott \etal 2008).\\

One of the free parameters in the method is the mass threshold \Mth~
used to make the reconstruction. In order to test for the influence of
the choice of this value in our results we have run two
reconstructions and analyses using two different mass thresholds, the
first one, \MthA~ corresponds to a halo mass of threshold
$10^{11.5}$\hMsun, while the second one is almost ten times larger,
\MthB=$10^{12.7}$\hMsun. Table \ref{tab:RecPars} summarizes all the
parameters used to run both reconstructions.

\begin{figure}
  \includegraphics[width=7.7cm,angle=270]{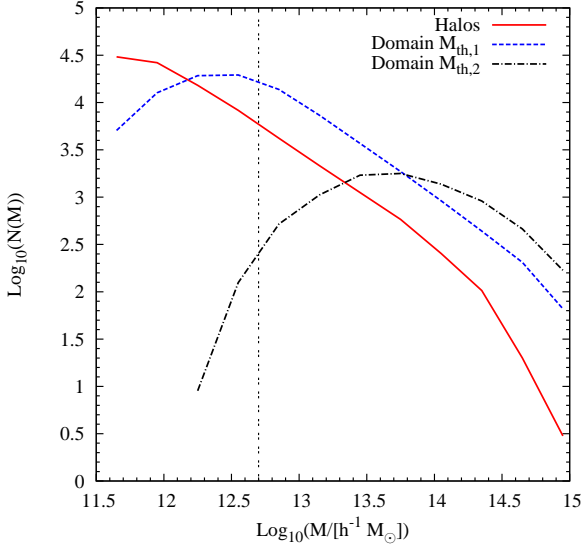}
  \caption{Mass functions for haloes and domains in both realizations
    \MthA~ and \MthB~ of the reconstructed density field. The red
    solid line shows the mass function for haloes in the halo catalog
    used to make the reconstruction while the blue dashed and black
    dot-dashed lines show the mass function for their domains in the
    reconstruction. The vertical line at $\mathrm{log}_{10}M=12.7$
    marks the mass threshold for \MthB~ in the halo mass function.}
  \label{fig:MfuncMdomMvir}
\end{figure}

Figure \ref{fig:MdomMvir} shows the initial and final mass assignment
in domains of haloes of mass $M$. The scatter plot shows the
distribution of masses for haloes in the realization with \MthA. The
step-like shape in the distribution of points is due to the
discreteness in the assignment of mass in domains for haloes in a
given population. The solid blue line shows a linear fit to the data
for \MthA. The blue dashed line shows the initial mass assigned to
domains of haloes of mass $M$ for that mass threshold. The difference
between the initial and final mass assigned is due to the overlapping
of the domains during the reconstruction that reduces the mass in
domains. For \MthA~ the power law relation between the final mass in
domains and $\Mvir$ has the same power law index as the one for the
initial mass assignment, implying that on average, for a given halo
mass, the final mass assigned in domains is a factor of $\sim10$ times
lower than initially assigned. For the reconstruction with \MthB, the
final \Mdom-$\Mvir$ relation has a slightly different power index,
this will imply a mass dependence between the ratio of the initial and
final mass in domains, giving more mass in domains to more massive
haloes. However the difference is very small, and the final mass in
domains in haloes in the reconstruction is around a factor of $\sim30$
times smaller than initially assigned.

Figure \ref{fig:MfuncMdomMvir} shows the mass function of haloes
(number of haloes with masses between $M$ and $M+dM$) and domains. The
red solid line shows the mass function for the halo catalog used in
the reconstruction, while the blue dashed and black dot-dashed lines
are the mass functions for domains in the reconstructions for
\MthA~and \MthB. The mass function for haloes is the same for both
realizations, only that for \MthB~ it is truncated at the low end at
$10^{12.7}$\hMsun, as shown by the vertical line in the Figure. It can
be seen that higher values of \Mth~ lead to higher masses assigned to
domains, as was expected from Figure \ref{fig:Massignement}. One can
see from the Figure that the mass function for domains is always
larger than the one for haloes except for masses below
$10^{12.2}$\hMsun~ for \MthA~ and $10^{13.3}$\hMsun~ for \MthB, where
the number of haloes with domains of mass $M$ decreases and even it
can be possible to have haloes with domains less massive than the halo
itself. Such a behavior is expected for haloes close to \Mth. Note
also that the mass functions for the domains cross each other at
$10^{13.8}$\hMsun~ and for a given mass, the number of haloes with
massive domain is higher for \MthB~ relative to \MthA. The higher
amplitude of the domain mass function for \MthB~ may be responsible
for the different slope observed in Figure \ref{fig:MdomMvir} for the
mean mass halo-domain relation. Note also the similarity of this plot
with the one shown in Figure \ref{fig:Mfunctions} where we show the
equivalent quantities for the haloes in the simulation.

A point worth to be verified in our implementation of the method is
the assumption of the correspondence of the density profiles computed
for haloes in the simulation in real space, with haloes in the survey
that are in redshift space. As pointed out in Wang \etal (2009), since
the method is based on individual haloes, each associated to a group
of galaxies, all the strong nonlinear redshift distortions are
absorbed in to the selection of individual haloes for the
reconstruction. Mildly nonlinear and linear redshift distortions
should still exist anyway in the halo catalog, and we expect them to
be not strong enough to affect the final reconstruction.
  
\begin{figure}
  \includegraphics[width=7.9cm,angle=270]{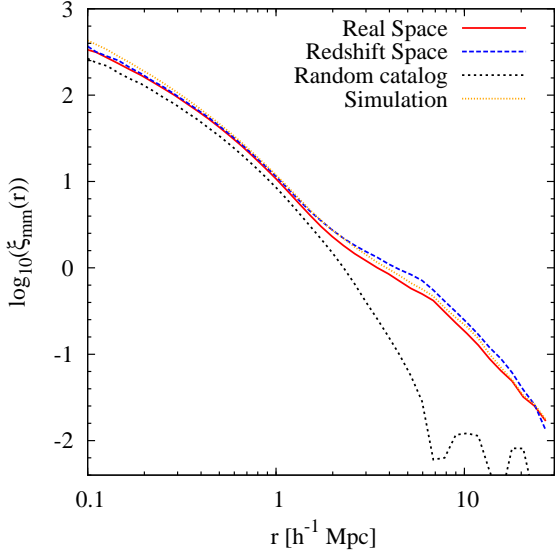}
  \caption{Correlation function computed for the reconstructed density
    field of the simulation in real and redshift space. Also are shown
    the correlation functions for the reconstruction using a random
    catalog, and as a comparison, the correlation function of the
    particle distribution in the original simulation. Error bars are
    not shown. Because the high number of points used to compute
    $\xi(r)$, they are smaller than the simbols.}
  \label{fig:correlationSim}
\end{figure}

To test this assumption, we compare two reconstructions of the mass
density field using the halo catalog taken from the cosmological
simulation described in section \ref{sec:sims}. One of the catalogs
has the original positions of the haloes in real space, while for the
other one we pick a halo close to the center of the simulation
box. Relative to this halo we recompute the positions of all other
haloes in redshift space according to their peculiar velocities,
mimicking the redshift space distortions. Then we build the
reconstructions using in both cases the same set of functions
$\eta(r,M)$, computed in real space. We compute the correlation
function for both reconstructions following the procedures shown in
section \ref{sec:corr} and find very small differences between the
reconstructions in real and redshift space. As can be seen in Figure
\ref{fig:correlationSim}, the effect of using the real space density
profiles makes the haloes to be slightly more clustered on scales
$>1.5$ \hMpc. The clustering pattern at scales larger than
$\sim1$\hMpc~ is dominated by the large scale clustering of haloes,
while for scales shorter than $\sim1$\hMpc~ the shape of the
correlation function is dominated by the typical mass distribution of
haloes in real space. We conclude after this test, that if well, it is
true that the use of the density profiles $\eta(r,M)$ computed in real
space is not fully consistent with its use on a halo catalog that is
built in redshift space. However the fact of the use of haloes as
centers to convolve the density distribution reduces the effect of the
redshift distortions to levels we consider acceptable.

It is important to note that the method can be implemented in real
space, as it has been shown in Wang \etal (2009). In the current work
we just focus on the redshift space reconstruction and reserve for
further papers the implementation of the method in real space.

\subsection{Mean mass density}

Another quantity that can be estimated from the reconstruction is the
mean mass density in the volume of the survey. We compute it in radial
comoving shells of width $dr$ centered in the center of the
survey. For this we use a procedure similar to the one used in section
\ref{sec:etar} to compute the value of $\eta(r,M)$ for each population
of haloes. First we associate a volume to each particle in the
reconstruction using a Delaunay tessellation, then we group the
particles in radial bins of width $dr$ and compute the density in that
bin as the sum of the mass of the particles inside the bin divided by
the total volume, the last, computed as the sum of the volumes of all
the particles inside that bin. The width for the radial bins is fixed
such that each bin contains the same spherical volume. We compute the
density distribution for the halo catalog using the same procedure but
in this case the mass in each spherical bin is the sum of the masses
of the haloes inside the bin.

\begin{figure}
  \includegraphics[width=7.7cm,angle=270]{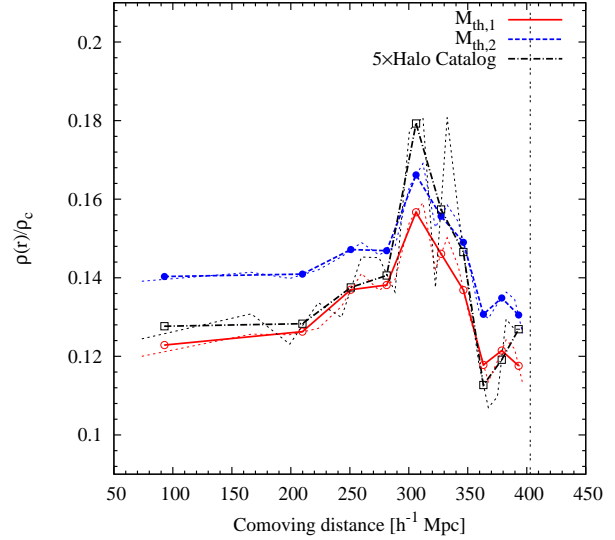}
  \caption{Mean mass density of the reconstruction as a function of
    the distance. The red line shows the mean density for the
    realization \MthA, the blue for \MthB~ and the black shows the
    mean mass density in haloes computed directly from the group
    catalog. Thick lines show the mean computed in 10 radial bins,
    thin lines show the calculation on 20 radial bins. The mean
    density of the halo catalog was scaled by a factor of 5 to scale
    it to the same order of the mean density in the
    reconstruction. The vertical line shows the outer border of the
    reconstruction.}
  \label{fig:Mean_density}
\end{figure}

Figure \ref{fig:Mean_density} shows the result of the density estimate
as a function of comoving distance. We used 10 and 20 radial bins and
no major differences were found for the values of $\rho(r)$ except for
the halo catalog where the peak initially seen at $\sim$300\hMpc~ is
resolved in two different peaks at $\sim$300 and $\sim$325 \hMpc. This
double peak is also seen in the density distribution of the two
reconstructions but the difference is not as drastic as for the halo
catalog. As one expects, the radial mass density changes with distance
but remains close to the mean value, being nearly constant for the
first $\sim$200\hMpc~ but presenting a peak at $\sim$310\hMpc. That
peak is associated with the presence of the Great Wall of galaxies
which makes the local mass density to increase by about 20\% relative
to the mean density. Beyond the Great Wall, the density drops below
the mean value again mainly because right behind the Great Wall is the
border of our reconstruction.

The total mean mass density in each case is $\bar{\rho}/\rho_c= $0.13,
0.14 and 0.03 for \MthA, \MthB~ and for the halo catalog respectively.
Clearly the mean mass density for the halo catalog is around a factor
of five times lower than for the (reference) reconstruction with
\MthA. These values for the mean density are off by factors of 1.8,
1.66 and 8.7 relative to the cosmological mean mass density of
$\Omega_m=$0.24. Although these results are acceptable for the mean
cosmic mass we computed in our reconstruction without the assumption
of any bias, we assume the difference between the reconstructions and
the expected values of the mean density by a factor of ~2 are due
mainly to the masses assigned to haloes in the halo catalog.

To verify this affirmation, and to test that the disagreement in the
values of the mean mass density in the reconstruction is not due to
the method or our implementation, we run a test using a cosmological
simulation of 1\hGpc~ volume. From the catalog of haloes in that
simulation we take a sample of haloes with the same geometry and
selection criteria we used to build the halo catalog from the group
catalog of the SDSS, but in this case assuming that the halo mass is
the one computed from the particle distribution in the
simulation. Then we ran the reconstruction using that set of haloes
and compute its mean density. As a result, we obtain for this test
reconstruction a mean density of $\bar{\rho}/\rho_c= 0.238$, in very
good agreement with the simulation input. This verifies that the
factor of around 2 difference in the mean mass density we see in our
reconstructions of the density field is not due to the reconstruction
method, but is likely related to the properties of the group catalog
and potentially due to the presence of cosmic variance.

\subsection{Correlation functions}
\label{sec:corr}
\begin{figure}
  \includegraphics[width=7.9cm,angle=270]{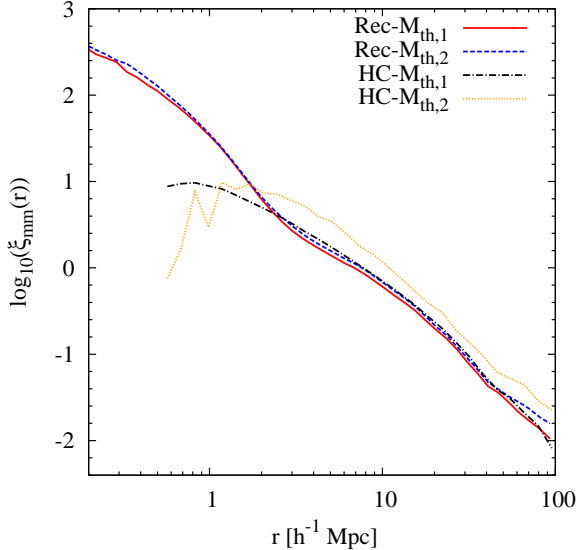}
  \caption{Correlation functions computed for the reconstructed
    density field (Rec) and for the group catalog (HC) in the two mass
    thresholds. Error bars are not shown. Because the high number of
    points used to compute $\xi(r)$, they are smaller than the
    simbols.}
  \label{fig:correlation}
\end{figure}

Figure \ref{fig:correlation} shows the mass correlation function
$\xi_{mm}(r)$ for the reconstructions with \MthA~ and \MthB~ and the
correlation function for the underlying halo catalog for the two mass
thresholds. To estimate the correlation functions in the
reconstructions we take from each of them random subsamples of data
points of $N_d \sim 3\times10^6$ particles. We use those samples to
compute the two point correlation function with an uniform random
catalog with equal number of random points. The selection of the size
of the sample of points $N_d$ should not affect the estimated
correlation function. This is clear if one thinks that reducing the
particle resolution of the reconstruction, it will end up consisting
on a lower number of particles (i.e. as small as $3\times10^6$ if one
chooses the appropriated value for $m_{\rm rec}$) that by construction
should follow the same density distribution as the reconstruction with
a high resolution, therefore, giving rise to the same correlation
function, only affected in the low scales by the low resolution of the
reconstruction. However, to test for the response of the estimates of
the correlation function to the arbitrariness of the number of points
of the subsample $N_d$, we repeate the procedure with $N_d=10^5$ and
$10^6$. We found no major dependence in the correlation function on
$N_d$. Figure \ref{fig:correlation} shows that the reconstruction
clearly exhibits the two terms of the correlation function (Zehavi
\etal 2004, Hayashi \& White 2008), the one-halo term dominating the
correlation function up to $r\sim2$ \hMpc~ and the halo-halo term
dominating for $r > 2$ \hMpc.

As it has been already shown (Wang \etal 2008, Hayashi \& White 2008)
more massive haloes are more strongly clustered. It can be seen in
Figure \ref{fig:correlation} for the correlation functions of haloes
in the halo catalog for \MthA~ and \MthB~ where $\xi(r)$ is large for
the most massive population of haloes \MthB. It is very interesting to
note that, independent on the biasing of the halo population used to
perform the reconstruction, the correlation function of the mass
distribution in the reconstruction is almost the same. We find that
our reconstructions of the cosmic mass density field are bias free in
the sense that although there is not explicit assumption about bias,
the reconstruction method converges for the mass clustering pattern
(mass correlation function) for different values of \Mth. This result
is also a proof of the consistence of the reconstruction, since it
approaches very well the clustering pattern of the underlying
cosmological model, as it is shown in Figures \ref{fig:correlationSim}
and \ref{fig:correlation}.

We have to explore the reason of the small deviation in $\xi(r)$ for
$r>30$\hMpc. In principle it can be due to fact that computing
$\eta(r,M)$ for a mass threshold as high as \MthB~ requires the use of
simulations with a larger volume than the one used in this work to
allow a better sampling of the environment of haloes on the very large
scales.

We try to identify the scales where the results of our reconstructions
are dominated by the priors of the method, specifically, by the shape
of the functions $\eta(r,M)$. We take the catalog of haloes from the
simulation, and replace the original positions of the haloes by random
positions uniformly distributed into the volume of the box. This
procedure should erase all physical relation between haloes in the box
while keeping the same mass function for the haloes in the
volume. Then we ran a reconstruction of the mass density field using
this artificial halo catalog and estimate the correlation function of
the reconstruction. After this experiment we observe that only the
one-halo term in the correlation function is dominant, as can be seen
in Fig. \ref{fig:correlationSim}. Comparing this correlation function
to the one obtained from the reconstructed density field for the
simulation, we conclude that the reconstruction method dominates in
the scales of $<2$~\hMpc. This result was expected, since there is no
information from the observations for these small scales. For scales
larger than $2$~\hMpc~ the method preserves the original clustering
pattern, which is indication that there is no major influence on the
result at these scales.

\begin{figure}
  \includegraphics[width=8.0cm,angle=270]{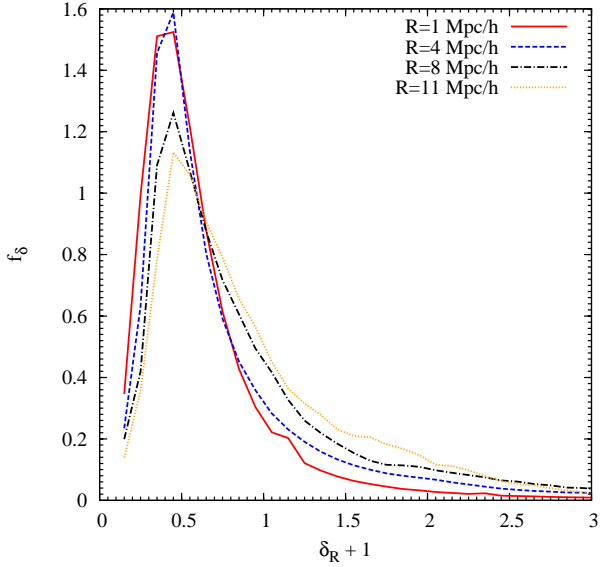}
  \caption{Distribution of counts in spherical regions of different
    radius computed for a sub box taken from the reconstruction with
    \MthA.}
  \label{fig:deltadists}
\end{figure}

\subsection{Statistics of counts in spheres}

Finally, we study the mass in the reconstruction through the particle
distribution, computed in spheres of radius $R$. In order to avoid the
inconvenients related with the geometry of the survey, we study the
count in cells for the mass distribution using a subsample volume. We
take two cubic volumes of side length 60 \hMpc~ from the north cap of
the survey and the reconstruction with \MthA, and compute the mass
distribution in spheres of radius $R$ from the particles of the
reconstruction. The subsamples containe $\sim2\times10^6$ particles
from the realization, and $N_{\rm rand}\sim10^7$ random points were
used to compute the counts in spheres. Figure \ref{fig:deltadists}
shows the distribution of the mean mass density contrast, $\delta_R$,
in spheres of radii $R$. The mean mass density contrast, $\delta_R$,
is defined as

\begin{equation}
  \delta_R = (\rho_s - \bar{\rho})/\bar{\rho},
\end{equation}

\noindent
where $\bar{\rho}$ is the mean mass density in the reconstruction and
$\rho_s$ is the mean mass density in the sphere. As expected, due to
the non-linearity of the scales studied, the distributions are not
Gaussian. However it can be seen that the distributions have a skewed
log-normal like shape, and broadens and moves to the right (tending to
1) for increasing values of $R$. This behavior is be expected (Coles
\& Jones 1991) if one assumes that going to larger values of $R$, the
mean density contrast should become more linear, therefore closer to
be Gaussian and having a mean $\bar{\delta}_R=0$

\begin{figure}
  \includegraphics[width=8.0cm,angle=270]{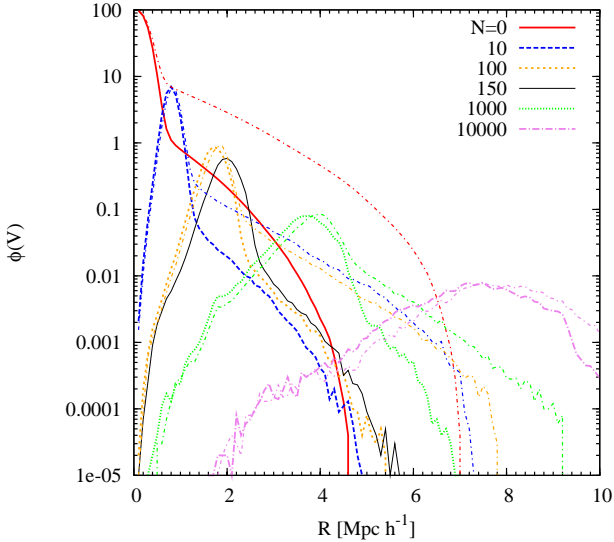}
  \caption{Distribution of the fraction (percent) of spheres of radius
    $R$ containing N particles. The mass enclosed in each sphere
    corresponds to 0, $8.9\times10^{10}$\hMsun,
    $8.9\times10^{11}$\hMsun, $8.9\times10^{12}$\hMsun~ and
    $8.9\times10^{13}$\hMsun~ respectively. The thin dot-dashed lines
    show $\phi_N(R)$ computed for the second subbox and show the
    effects of the cosmic variance on $\phi_N(R)$, while the solid
    black line is $\phi_N(R)$ for a mass in the sphere equal to
    $M_*$.}
  \label{fig:PofV}
\end{figure}

Following Maurogordato \& Lachieze-Rey (1987) we computed the number
of particles $N_i(R)$ inside all of the $N_{rand}$ random spheres
drawn inside the cubic box of 60\hMpc~ side. Then we compute the
probability of a sphere of radius $R$ to contain $N$ particles,
$\phi_N(R)$, as

\begin{equation}
  \phi_N(R) = \frac{N^{acum}_N(R)}{N_{rand}},
\end{equation}

\noindent
where

\begin{equation}
  N^{acum}_N(R) = \sum^{N_{rand}}_{i=1} [N_i(R) = N],
\end{equation}

\noindent
is the sum of all the spheres containing $N$ particles. Figure
\ref{fig:PofV} shows the fraction of spheres of radius $R$ containing
$N$ particles for $N=0, 10, 100, 1000$ and $10000$. Note that this
number of particles can be translated to mass, so, Figure
\ref{fig:PofV} shows the fraction of spheres of radius $R$ containing
masses of 0, $8.9\times10^{10}$, $8.9\times10^{11}$,
$8.9\times10^{12}$ and $8.9\times10^{13}$\hMsun~ respectively. To be
able to compare with the mean density predicted by the cosmological
model, we have accounted for the fact that the reconstruction has a
mass assignment biased by a factor of $\sim1.8$ relative to the true
cosmic mass, as was shown in the previous sections. To test for the
effects of the cosmic variance introduced by the small size of the
boxes used (60\hMpc), we plot in thin lines (thin dot-dashed lines)
the same quantity, $\phi_N(R)$, for a second subbox of the same
size. It can be clearly seen that for both boxes the shape of the
distribution is the same, the peaks are located nearly at the same
radius but they are broader for the second subbox. These differences
are simply due to the second subbox to be taken from a region with a
lower mean density than the first one.

As it can be seen from the Figure \ref{fig:PofV}, requiring the
spheres to be totally empty forces the distribution to be peaked at
$R=0$, as one would expect in that case it will be more probable for
the smaller spheres to be empty. Requiring spheres to contain
$8.9\times10^{10}$\hMsun~ shows that it is more probable to find them
having a radius of $R=0.7$\hMpc~ and spheres containing
$8.9\times10^{11}$\hMsun~ will be found more frequently in spheres of
radius $R\sim1.7$\hMpc. Of special interest is to see what is the most
probable radius of the sphere containing a mass of
$\sim1.3\times10^{12}$\hMsun~ (solid black line in Figure
\ref{fig:PofV}). As it can be seen in the Figure, that amount of mass
can be found most probably in spheres of radius $R\sim1.8$\hMpc. This
result is another independent verification of the ability of our
reconstruction to fit the expectations from the models, since for
$z=0$ and the cosmology assumed for this work, the characteristic
non-linear mass is $M_* \sim 1.5\times10^{12}$\hMsun, and the related
radius $R_*=1.7$\hMpc, in very good agreement with the previous
result. This also supports the fact that the mismatch in the mean mass
densities shown in the previous sections is mostly due to the mass
assumed for the haloes from the group catalog.

\begin{figure*}
  \includegraphics[width=14.0cm,angle=270]{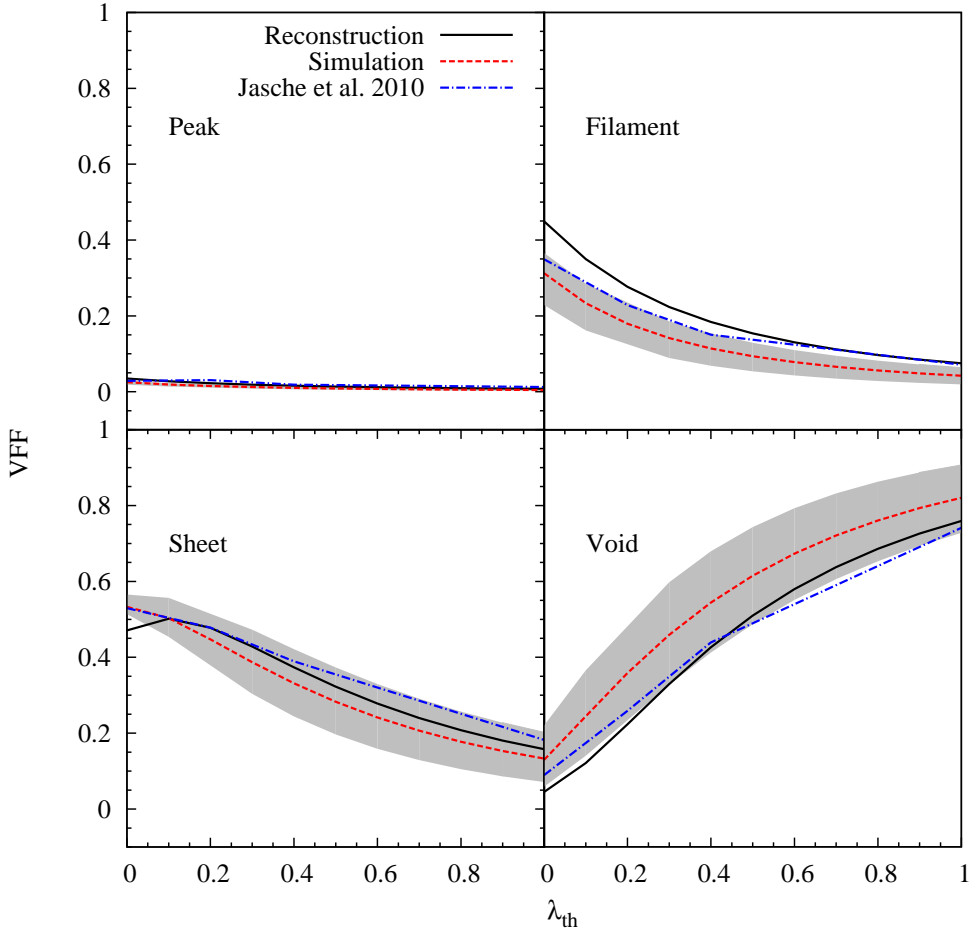}
  \caption{Volume filling fractions associated to each environment
    characteristic as a function of $\lambda_{th}$ computed for the
    density field taken from the reconstruction and for a
    simulation. Gray region shows the cosmic variance estimated in
    Forero-Romero \etal (2009), the red dashed line shows the
    estimation for the cosmological simulation, the black solid line
    the estimations for our reconstruction, while the blue dot-dashed
    lines show the data from Jasche \etal (2010).}
  \label{fig:VFF}
\end{figure*}

\subsection{Characterization of environments}

We now investigate the morphological classification of the cosmic web
in our reconstruction of the density field. Following Forero-Romero
\etal (2009) we classify the different environments of the
reconstructed density field in peaks, filaments, sheets and voids,
according to the the eigenvalues of the deformation tensor of the
cosmic mass distribution. To this end, we compute the smoothed density
field arising from the particle distribution on a regular cubic grid
of size 1\hGpc~ and $512^3$ cells, and fill the empty regions of the
survey with a set of particles randomly distributed with a mean mass
density equal to the mean mass density of the reconstruction. Then we
compute the potential of the smoothed density field in the grid and
from it the deformation tensor

\begin{equation}
T_{\alpha\beta}=\frac{\partial^2\phi}{\partial x_{\alpha}\partial x_{\beta}}.
\end{equation}

\noindent
We classify the environment computing the eigenvalues of the
deformation tensor for every cell and compare them with a threshold
$\lambda_{th}$ as a confidence value for the choice of the
classification. The type of environment assigned to each cell is
determined by the number of eigenvalues above $\lambda_{th}$. Such a
procedure will allow us to study in detail not only the geometrical
properties of the mass distribution through the cosmic web, but it
will also provide a direct link between environments in the volume of
the reconstruction and the properties of the galaxies residing in
these environments.

Figure \ref{fig:VFF} shows the volume filling fractions (VFF) for each
environment characteristic computed for the reconstructed density
field and for a cosmological simulation, for a comparison. For the
simulation we used the 1\hGpc~ simulation already mentioned. In the
volume of that simulation we took a piece with the same geometry of
the survey and produced an equivalent background distribution. As can
be seen in Figure \ref{fig:VFF}, the volume filling fractions for peak
and sheet environments in the reconstruction follows very well the
same behavior observed for the simulation. Differences are observed
for filaments and voids. As it has been shown in Forero-Romero \etal
(2009), the differences may be due to fluctuations induced by cosmic
variance. One can verify this by comparing our VFF with the ones
obtained by Jasche \etal (2010) for the reconstructed density field of
the SDSS-DR7 using a completely different approach. The agreement for
both estimates of the VFF for roughly the same region of the universe
indicates that the observed differences relative to the simulation are
mostly due to cosmic variance.

Note that the volume filling fractions for filaments, peaks and sheets
decrease with the increase of $\lambda_{th}$. That behavior is be
expected since increasing the value of $\lambda_{th}$ decreases the
probability for the eigenvalues of $T_{\alpha\beta}$ to be larger than
the given $\lambda_{th}$, then increasing the number of cells
classified as voids and necessarily decreasing the number of cells
classified as the other environments.

\begin{figure*}
  \includegraphics[width=5.8cm,angle=0]{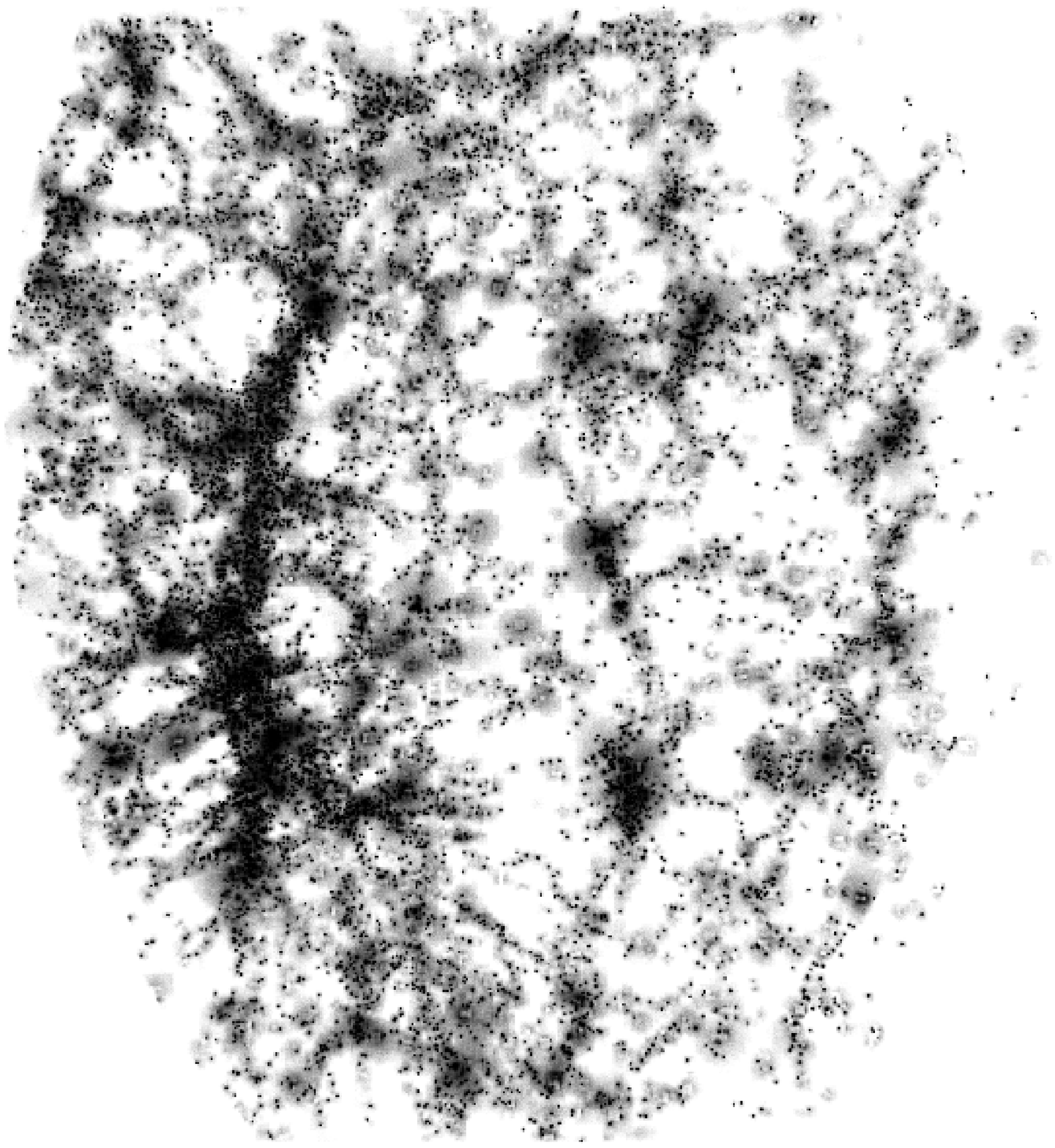}
  \includegraphics[width=5.8cm,angle=0]{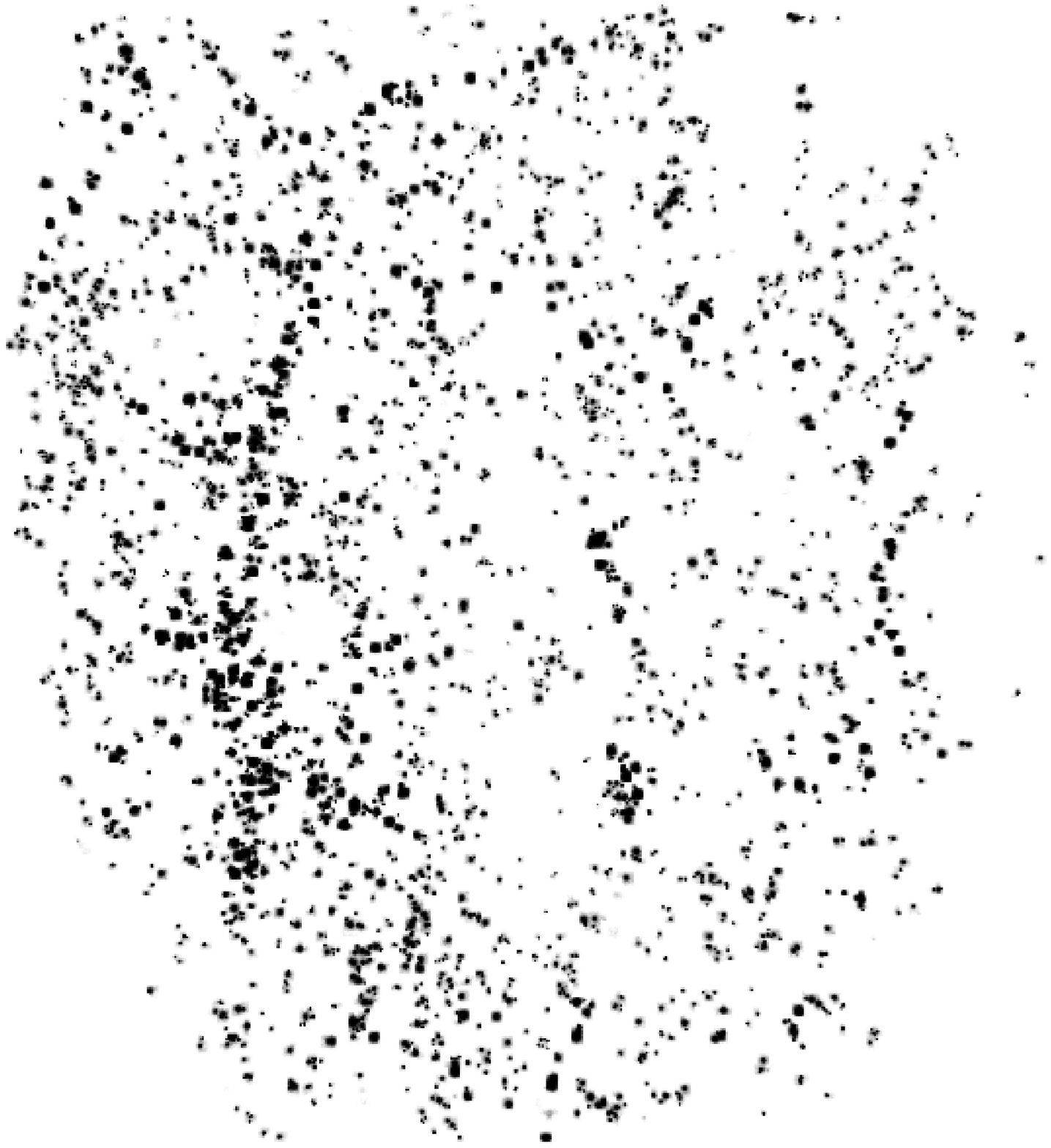}\\
  \includegraphics[width=5.8cm,angle=0]{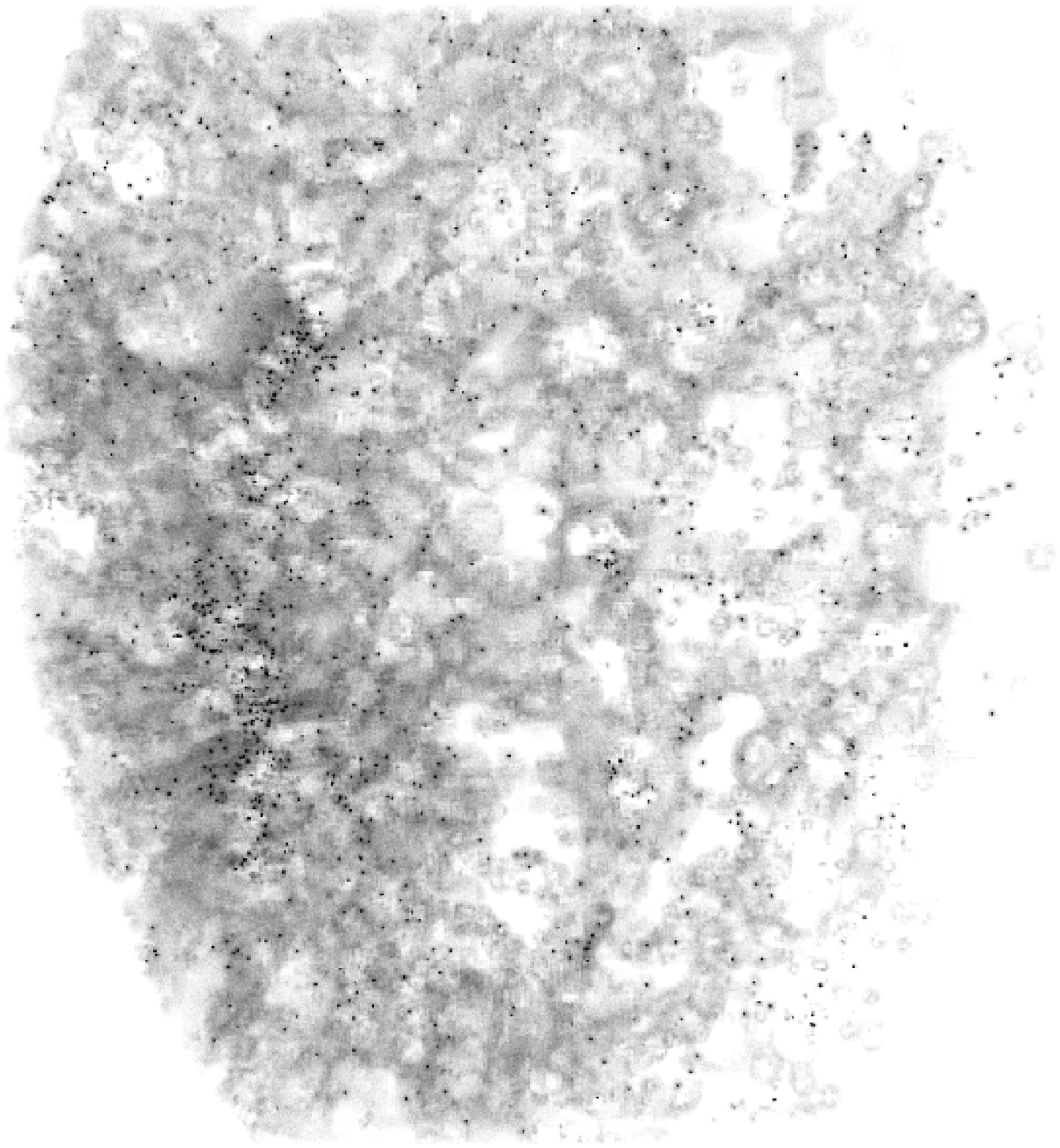}
  \includegraphics[width=5.8cm,angle=0]{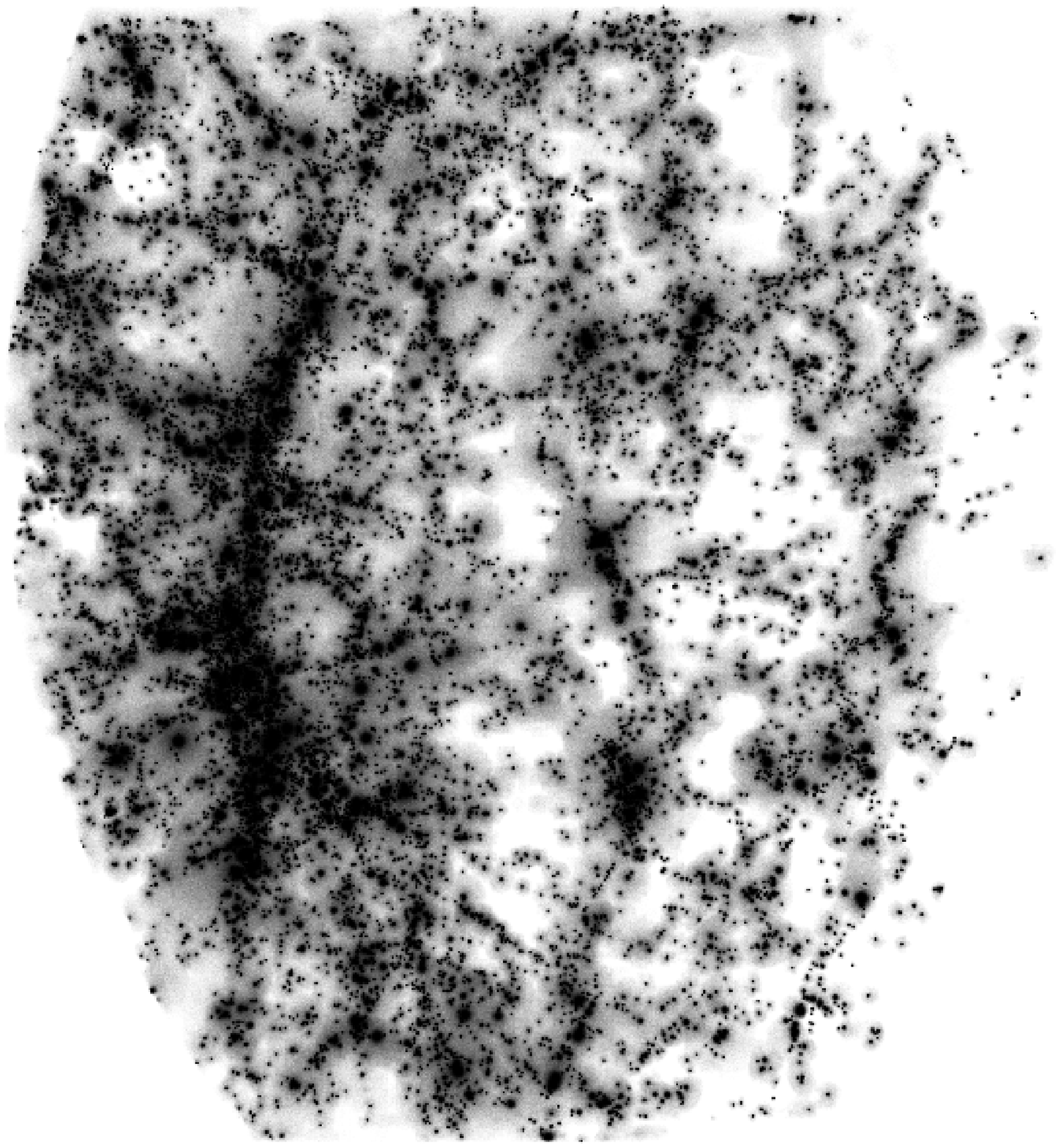}
  \caption{Classification of the cosmic web in a region enclosing the
    Great Wall using $\lambda_{th}=0.2$. From the top left to the
    bottom right, the panels show individually the regions classified
    as filaments, peaks, sheets and finally, all of them together.}
  \label{fig:Env}
\end{figure*}

Figure \ref{fig:Env} presents a visual impression of the results of
the classification of the cosmic network in our reconstruction for
$\lambda_{th}=0.2$ (see Forero-Romero \etal 2009 for details in the
selection of this value). The Figure shows a piece of the survey of
$400 \times 400$ \hMpc~ and $\sim30$\hMpc~ thick, enclosing the Sloan
Great Wall. The panels show how the classification recognizes
filaments, peaks and sheets in the reconstruction. It can be clearly
seen how the peaks are distributed following the structure of the
filaments. Note also that although there are peaks inside the regions
classified as filaments, which is partly due to the discreteness of
the grid used to compute the eigenvalues of $T_{\alpha\beta}$, but
also due to the inherent nature of haloes to be located in filaments,
the filaments are characterized by the smoothly extended mass
associated to the environment of the haloes. The last panel shows
together the regions classified as peak and filaments, and it can be
clearly seen where are the empty regions (voids) surrounded by
filaments.

\section{Discussion and Conclusions}

We have presented the implementation of a halo based reconstruction of
the cosmic mass density field. The method combines the results of
observations and simulations, convolving the mass density distribution
in and around dark matter haloes computed from N-body simulations with
the coordinates of haloes of a given mass identified in the group
catalog of a galaxy survey. We have used the group catalog of Yang
\etal (2007) built from the SDSS-DR4 together with high resolution
simulations of the formation of the structure in the standard
spatially flat $\Lambda$CDM model.

The reconstruction method produces a distribution of points sampling
the cosmic mass density field that is determined by the mass and
position of the dark matter haloes in the halo catalog, convolved with
the estimated typical density distribution of mass in haloes and their
environments. The final particle distribution resembles that obtained
from cosmological simulations, and as the reconstruction method does
not requires the definition of any scale length, it provides a high
dynamical resolution. Tracer particles are more probably found in high
density regions, however the adaptivity of the method allows to sample
also the low density regions of the cosmic web.

Instead of imposing a smoothing length, the resolution and smoothness
of the method is limited by two factors, the value of the mass
threshold \Mth~ and the mass of the sampling particles
$m_{\mathrm{rec}}$. The resolution of the reconstruction is set by the
choice of the mass threshold \Mth. Using lower values for the mass
threshold produce a more detailed reconstruction of the density field
at small scales. The selection of the mass threshold plays a role
similar to the smoothing length in standard grid techniques of
computing the density field, like cloud in cell (CIC), where
structures below the threshold value (smaller than the smoothing
length in the case of the CIC method) are erased and smoothed out in
the density field. A key result is that, as was shown throughout the
paper, the choice of the value of the mass threshold does not affect
the global properties of the reconstructed density field.

The other parameter defining the the smoothness of the reconstruction
is the mass of the particles used to sample the density field,
$m_{\mathrm{rec}}$. Low mass sampling particles produce a smooth and
more well defined sampling of the density field, specifically in the
case of haloes with masses close to the mass threshold \Mth. In
general $m_{\mathrm{rec}}$ has to be lower than \Mth, to allow the
particle distribution to reliably reproduce the mass associated with
haloes of masses close to the mass threshold.

It is worth to note that the reconstruction method is based on the
concept of individual haloes, that is, main haloes in the language
used in cosmological simulations. Due to observational constraints,
this method would be better to be used in populations of haloes in a
mass range above \Mth$ > 10^{11}$\hMsun, where is it is possible to
reliably identify complete samples galaxies and groups of galaxies
associated to haloes down to this mass value.

As it has been shown in section \ref{sec:results}, the shape of the
correlation function for our reconstructions using two different
values of \Mth~ converges to similar values at almost all scales.
From this result and the fact that the method does not make any
explicit assumption about biasing, the resulting reconstructions are a
very suitable tool to address studies of galaxy properties and their
relation with the large scale density field. The known effect of
biasing, for haloes of different mass is introduced in a natural way
in the reconstruction through the functions $\eta(r,M)$.

A drawback of the method (and of any reconstruction method in general)
is that it is strongly dependent on the properties of the catalog of
haloes as well as on the details of the typical mass distribution in
and around of haloes $\eta(r,M)$ extracted from the simulations. The
quality of the observations, and in particular, the properties of the
halo catalog, have an influence on the quality of the
reconstruction. It was shown that the mass assigned to haloes in the
halo catalog gives values of the mean cosmic density that are off by a
factor of around 2. Clearly better estimates of the masses of haloes
may produce a better constraint. Despite the discrepancy, we see this
result as motivating and as a good starting point in the improvement
on the methods used to assign masses in the halo catalog like the one
presented by Yang \etal (2007). On the other hand, as we have shown,
we can assume the masses in the halo catalog to be off by the same
factor of $\sim2$ as the density. This value is in complete agreement
with the measured scatter for the mass assigned to haloes in the group
catalog (Yang \etal 2008, Wang \etal 2008). We believe that the volume
of the survey have no influence on this mismatch since it is totally
independent on any assumption besides the cosmological parameters.

It is also true that assuming the typical density distribution of mass
from simulations will make the method not appropriate to extract
further cosmological information from the mass distribution in the
halo catalog. Such an assumption is similar to assuming a power
spectrum shape, as done in methods like the Wiener filter
techniques. As it has been already mentioned, the power of the method
relies on its physical simplicity, directly connected to the ideas of
the halo model, and it is best suited to make studies of the
properties of the mass distribution: large scale environments and
environmental effects on galaxy formations, and problems where there
is a pre-stablished cosmological model, like the realization of
constrained simulations.

Besides the aforementioned sources of error coming from the
observations, we have shown different potential sources of error in
our reconstruction, and we have shown where the reconstruction is more
strongly dominated by the data or by the priors of the
method. Nevertheless is difficult to quantify the degree of error
introduced in the reconstruction. We have estimated the amount of
deviation between the expected and reconstructed mean mass
density. This number may be taken partially as an estimation for the
mean error in the reconstruction.  Analysis in the clustering pattern
give also information about the degree of effects introduced by the
method at different scales. But in general, as far as it has been
shown in this and in previous implementations of the method, there is
not yet a formal model to properly estimate the errors of the
reconstructed density field in a given point, and will be part of
further research to study in deep this issue.

To check for the consistency of our results we have made an exhaustive
study of the properties of the reconstructed density field. We have
shown that the reconstructed density field reproduces the mass
correlation function. It clearly shows the two regions associated to
the one halo and two halo terms. Comparing the correlation function
for our reconstruction with the one for the halo catalog, we see that
the amplitude of the correlation function is in agreement with the
expected clustering at large scales, and by construction it reproduces
the small scale clustering. More interesting is that it shows the
bias-free nature of the method in the mass correlation function. We
computed the properties of the mass distribution in spheres and
verified the way the mass distribution follows a skewed, log-normal
like distribution when computed at small scales. As expected (Coles \&
Jones 1991), this distribution broadens and shifts to
$\bar{\delta}~\sim 0$ for larger smoothing radii.

The classification of the cosmic web for the reconstructed density
field also offered consistent results with the theoretical
expectations. The volume filling fraction of the different
characteristic environments (filaments, sheets, peaks and voids) in
the reconstruction match the ones obtained from simulations. Only
small differences within the expected cosmic variance are observed
(Forero-Romero \etal 2009).

Reconstructions of the density field as the one we have shown provide
a way to map directly the cosmic mass distribution from the halo
catalogs identified in the surveys without the necessity to assume any
functional form for the bias factor. Provided that the data fulfills a
minimum completeness requirements, coupling this method of
reconstruction of the density field with a reconstruction of the
velocity field will offer the opportunity of computing real space
density fields. This will give us detailed information about the
velocity field of the observed volume, information valuable to
quantify and map the effects of the redshift space distortion, and
usable to test the dynamics of the local neighborhood (Lavaux
\etal(2010a), Strauss \& Willick 1995).

Furthermore, the results of the cosmic web classification obtained
with the help of the reconstructed density field can be used to study
the properties of the galaxies in the large scale environment, not
only in their close environment (Tempel \etal 2010, Weinmann \etal
2009, 2008, Wang \etal 2008, Park \etal 2007).

The ability of the method to trace the mass distribution in an
adaptive way, coupled with the ability to work in real space, makes
the method a good choice to produce appropriate and high resolution
density fields to be used as background for time machines used in the
realization of constrained simulations structure formation (Kolatt
\etal 1996, Martinez-Vaquero \etal 2009, Gottl\"{o}ber \etal 2010,
Lavaux \etal 2010b, Nuza \etal 2010).

\section*{Acknowledgments}

J.C.M. was supported by the German Science Foundation under the grant
MU-1020/6-4. We thank to Sebastian Nuza and Stefan Gottl\"{o}ber for
useful comments to the manuscript, and to the referee for his/her
comments that helped to improve the presentation of this paper. Part
of the reconstruction and the analysis has been performed in the Altix
Supercomputer of the Leibniz Rechenzentrum (LRZ) Munich. The
simulation of the box 1\hGpc~ used for our tests has been performed by
A. Klypin and S. Gottl\"{o}ber at LRZ Munich within the German
AstroGrid-D project.

\bsp

\label{lastpage}

\end{document}